\documentclass[11pt,a4paper]{article}
\pdfoutput=1 

\usepackage[table]{xcolor}
\usepackage{colortbl}
\RequirePackage{ifpdf} 
\usepackage{amsmath} 
\usepackage{mathtools}
\usepackage{longtable}
\usepackage{jheppub}
\usepackage{pstricks}
\usepackage[final]{pdfpages} 
\usepackage{ifpdf} 
\usepackage{slashed}
\usepackage{numprint}
\usepackage[normalem]{ulem}
\usepackage{color}
\usepackage{xcolor}
\definecolor{urlblue}{rgb}{0.2,0.4,0.7}
\definecolor{citegreen}{rgb}{0,0.6,0.2}
\definecolor{linkred}{rgb}{0.9,0.2,0.1}
\usepackage{hyperref}
\hypersetup{
colorlinks=true, citecolor=citegreen, linkcolor=blue, urlcolor=urlblue}

\usepackage{graphics}
\usepackage{etoolbox} 
\usepackage{fixmath}
\usepackage{psfrag}
\usepackage{multido}
\usepackage[
  round-mode = places,
  round-precision = 2,
  round-pad = false 
]{siunitx}

\usepackage{notoccite} 

\usepackage{amsfonts}
\usepackage{autobreak}
\usepackage{marginnote}
\usepackage{enumitem}
\usepackage{appendix}
\usepackage{caption} 
\usepackage{multirow}
 \usepackage{pbox}
 \usepackage{array}
 \usepackage{float}
 \usepackage{lineno}
\newcolumntype{P}[1]{>{\centering\arraybackslash}p{#1}}
\newcommand{\NOdisplay}[1]{ }
\usepackage{pifont}

\usepackage{tikz}
\usetikzlibrary{positioning,arrows}
\usetikzlibrary{decorations.pathmorphing}
\usetikzlibrary{decorations.markings}
\usetikzlibrary{shapes.geometric}
\usepackage{endnotes}
\tikzset{
    vector/.style={decorate, decoration={snake}, draw},
    provector/.style={decorate, decoration={snake,amplitude=2.5pt}, draw},
    antivector/.style={decorate, decoration={snake,amplitude=-2.5pt}, draw},
    fermion/.style={draw=black,
      postaction={decorate},decoration={markings,mark=at position .55
        with {\arrow[draw=black]{>}}}},
    fermionbar/.style={draw=black, postaction={decorate},
                       decoration={markings,mark=at position .55 with {\arrow[draw=black]{<}}}},
    fermionnoarrow/.style={draw=black},
    gluon/.style={decorate, draw=black,decoration={coil,amplitude=4pt, segment length=6pt}},
    scalar/.style={dashed,draw=black,
      postaction={decorate},decoration={markings,mark=at position .55
        with {\arrow[draw=black]{>}}}},
    scalarbar/.style={dashed,draw=black,
      postaction={decorate},decoration={markings,mark=at position .55
        with {\arrow[draw=black]{<}}}},
    scalarnoarrow/.style={dashed,draw=black},
    electron/.style={draw=black,
      postaction={decorate},decoration={markings,mark=at position .55
        with {\arrow[draw=black]{>}}}},
    bigvector/.style={decorate, decoration={snake,amplitude=4pt}, draw},
}
\def\beq{\begin{equation}}
\def\eeq{\end{equation}}
\def\bsp#1\esp{\begin{split}#1\end{split}}
\newcommand{\rd}{\textrm{d}}
\newcommand{\claudecomment}[1]{\textcolor{red}{\bf [CD: #1]}}

\usepackage{siunitx}
\title{A comprehensive analysis of Drell-Yan production uncertainties and mass effects at moderate and low dilepton masses}

\author{Ekta Chaubey$^{a}$, Claude Duhr$^{a}$, Rhorry Gauld$^{b}$, Pooja Mukherjee$^{c}$}
\emailAdd{eekta@uni-bonn.de}
\emailAdd{cduhr@uni-bonn.de}
\affiliation{$^{a}$Bethe Center for Theoretical Physics, Universität Bonn,  53115 Bonn, Germany
}
\emailAdd{rgauld@mpp.mpg.de}
\affiliation{$^b$Max-Planck-Institut für Physik, Boltzmannstraße 8, 85748 Garching, Germany}
\affiliation{
$^c$II. Institut für Theoretische Physik, Universität Hamburg, Luruper Chaussee 149, 22761, Hamburg,
Germany}
\emailAdd{pooja.mukherjee@desy.de}

\preprint{BONN-TH-2025-26, DESY-25-096, MPP-2025-153}

\abstract{We present a thorough investigation of the sources of uncertainties to the Drell-Yan production using state-of-the-art predictions for both neutral and charged current channels, focusing on the low invariant mass region. Differential predictions for the invariant mass spectrum are provided at N$^3$LO supplemented with exact charm and bottom quark mass effects calculated at $\mathcal{O}(\alpha_s^2)$. The impact of PDF choices (including approximate N$^3$LO), scale variations, the variation of the strong coupling constant, and impact heavy quark mass effects on the distributions is studied in detail. We also comment on the correlation of high-energy astrophysical processes with the low-mass DY region.
}

\keywords{$Z$ production, $W$ production, QCD, VFNS.}

\begin{document}
\allowdisplaybreaks[4]
\unitlength1cm
\keywords{}
\maketitle
\flushbottom
\newcommand{\EC}[1]{\textcolor{red}{EC: #1}}
\newcommand{\RG}[1]{\textcolor{orange}{RG: #1}}


\section{Introduction}
\label{sec:intro}

High-energy hadron-hadron colliders provide unique opportunities to improve our understanding of fundamental physics. These opportunities include searches for signals beyond the Standard Model (SM) as well as precision measurements of known SM processes. Measurements of these kinds provide important information on known SM parameters, invaluable knowledge of hadron structure, and furthermore offer the possibility to test the validity of factorisation theorems which are employed to deliver precise SM predictions.

A key SM process at hadron-hadron colliders is the Drell-Yan (DY) process, which serves as a mechanism for lepton pair production and proceeds via either charged-current (CC) or neutral-current (NC) gauge boson exchange. Measurements of the DY processes have been performed at numerous hadron-hadron colliders (ranging in hadronic center-of-mass energies from the GeV to TeV range) over the past decades, which have in turn provided vast amounts of information on this scattering process in different kinematic regimes. The Large Hadron Collider (LHC) has further enabled measurements of this process, which have been performed with unprecedented precision. With the LHC set to enter a high-luminosity phase in the coming years, it is expected that measurements of this benchmark process will continue to improve.

This success of the experiments, in turn, places demands on the theoretical community to advance the precision of the corresponding theory predictions. 
In several cases, such theoretical advancements have made progress.
Since the original work of Drell and Yan~\cite{Drell:1970wh}, next-to-leading order (NLO)~\cite{Altarelli:1979ub} and next-to-next-to-leading order  (NNLO)~\cite{Hamberg:1990np,Anastasiou:2003ds,Melnikov:2006kv,Catani:2009sm,Catani:2010en} QCD corrections for the DY process have been known. More recently, the calculation of the DY cross-section has been made available up to third order (N$^3$LO) in perturbative QCD for the inclusive cross-section~\cite{Duhr:2020seh,Duhr:2020sdp,Duhr:2021vwj,Baglio:2022wzu} and predictions at the differential level have also been presented in~\cite{Chen:2021vtu,Chen:2022cgv,Chen:2022lwc,Neumann:2022lft,Campbell:2023lcy}. 
%
For differential observables, electroweak~\cite{Dittmaier:2001ay,Baur:2004ig,CarloniCalame:2007cd} (EW) corrections as well as mixed QCD-EW~\cite{Dittmaier:2015rxo,Dittmaier:2020vra,Behring:2020cqi,Buonocore:2021rxx,Behring:2021adr,Dittmaier:2024row} corrections have also been performed, and results for quantifying the impact of finite quark mass effects have been presented in~\cite{Gauld:2021zmq}.
%
Advancements in the extraction of parton distribution functions (PDFs) (and the required splitting functions) have also begun to reach a new level of precision, with partial results available for the four-loop splitting functions~\cite{Basdew-Sharma:2022vya,Falcioni:2023luc,Falcioni:2023vqq,Falcioni:2023tzp,Falcioni:2024xyt,Falcioni:2024qpd,Gehrmann:2023cqm, Gehrmann:2023iah,Kniehl:2025ttz} and global analyses of PDFs at approximate N$^3$LO accuracy~\cite{McGowan:2022nag, NNPDF:2024nan}.
It is therefore reasonable to expect that comparisons of DY data with differential N$^3$LO QCD predictions will become standard in the future.
Such a standard would offer new opportunities for global analyses of PDFs, where interesting kinematic regimes can be studied more reliably. 

\begin{figure}[!htbp]
    \centering
    \includegraphics[width=0.49\linewidth]{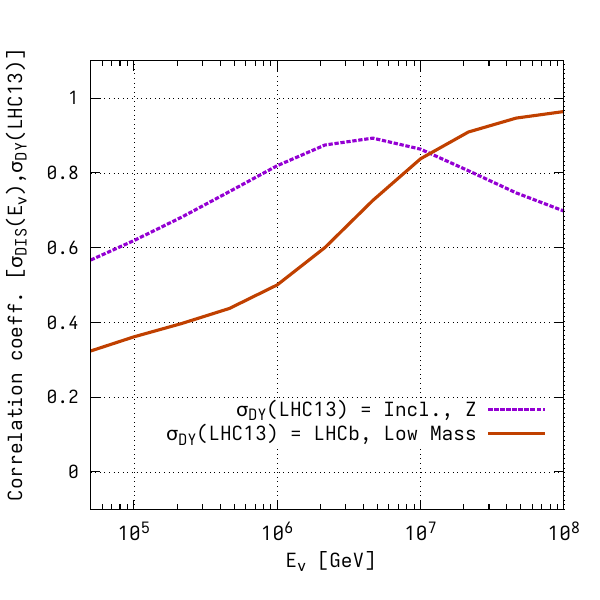}
    \includegraphics[width=0.49\linewidth]{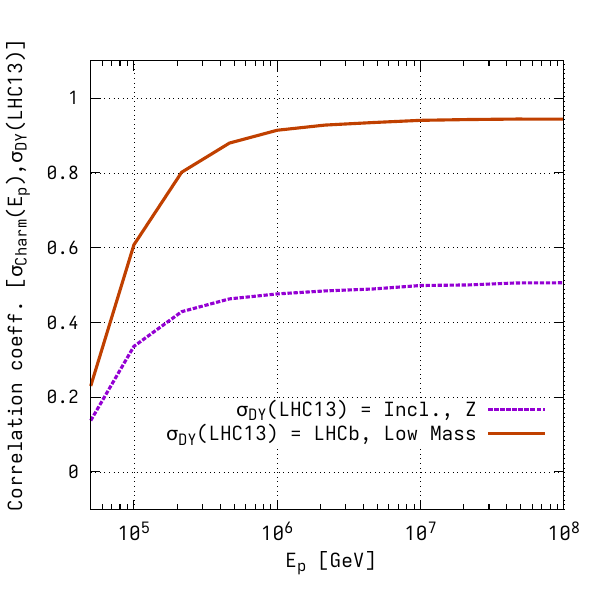}    
    \caption{Correlation coefficient between the cross-section rates for the neutral-current DY process at the LHC with ultra-high-energy neutrino-nucleon DIS (left) or inclusive charm production in hadron-hadron collisions (right).  In both cases the energy of the projectile neutrino $E_{\nu}$ (left) or proton (right) is varied while the target hadron is considered at rest.
     \label{fig:correlation}}
\end{figure}

A particularly interesting kinematic regime that can be accessed at the LHC is the region of small invariant dilepton mass ($m_{\ell \bar \ell}=Q$), and in addition forward rapidity, as made accessible by the LHCb experiment. For example, at leading order (LO), the production of a NC dilepton pair at rapidity $y$ in hadronic collisions with a center-of-mass energy of $\sqrt{S}$ offers sensitivity to PDFs carrying momentum fractions $x_{1(2)} = Q/\sqrt{S}\, e^{+(-) y}\,$.
Thus, measurements at relatively low dilepton invariant masses and large forward (backward) rapidity can provide information on hadron structure at values of small $x$ and at small virtualities that are currently not strongly constrained by data.
An example of a measurement of this kind has been presented by ATLAS~\cite{ATLAS:2014ape}, which is available for $Q \gtrsim 14 {\rm~GeV}$ and central rapidity.
A similar measurement is underway at the LHCb experiment~\cite{Plews:2022pmw}, where for values of $Q\sim 10{\rm~GeV}$, $y \sim 5$ and $\sqrt{S} \sim 13{\rm~TeV}$ the region of $x \sim 10^{-5}$ becomes accessible.
Knowledge of hadron structure in this region has important consequences for neutrino astronomy: It is relevant for the theoretical prediction of Ultra-High-Energy neutrino-nucleon scattering~\cite{Cooper-Sarkar:2011jtt,Bertone:2018dse,AbdulKhalek:2022fyi,Xie:2023suk}, which plays a role in interpreting measurements of neutrinos detected at large volume detectors. It is also an important ingredient for predicting (background) rates of atmospheric neutrinos produced by collisions of cosmic rays with Earth's upper atmosphere~\cite{Bhattacharya:2015jpa,Garzelli:2015psa,Gauld:2015kvh,Benzke:2017yjn,Zenaiev:2019ktw,Bai:2022xad}. 
To better quantify these statements in fig.~\ref{fig:correlation} we show the correlation coefficient between the DY cross-section rate at the LHC (at a hadronic center-of-mass energy of 13~TeV) with either the neutrino-nucleon cross-section (left) or the total charm quark production rate in proton-proton collisions (right).
The reference cross-section for the NCDY process is calculated for the inclusive rate with $Q \sim m_{Z}$ (labeled as ``Incl., Z") or for an invariant mass in the range $Q \in [10,15]~{\rm GeV}$ in the forward rapidity region $y_{\ell\bar\ell} \in [3.5,4.5]$ (labeled as ``LHCb, Low Mass"). 
The correlation coefficient, see eq.~(25) of~\cite{Alekhin:2011sk}, is then calculated with respect to either the neutrino-induced charged-current DIS cross-section as a function of increasing $E_{\nu}$ (in the nucleon rest frame) or the inclusive charm cross-section in proton-proton collisions as a function of one of the incoming protons energy (labeled as ``$E_{\rm p}$"), while the other is considered at rest.
These predictions have been calculated at NLO QCD accuracy with the PDF set {\tt {NNPDF40\_nlo\_as\_01180}}~\cite{NNPDF:2021njg}.
These figures demonstrate the important correlation between the DY process at small invariant masses (primarily sensitivity to PDFs at small $x$) and energetic processes relevant for neutrino astronomy.
The availability of measurements of the low-mass NCDY process at the LHC can be important from this point of view.

However, it is important to bear in mind that the corresponding theoretical prediction for the NCDY process is challenging at low invariant mass. The prediction must be performed at an energy scale $Q$ in which the convergence of the perturbative expansion of the cross-section is expected to be slower, and it is also unclear to what extent other physics effects (such as the role of quark mass effects) can be relevant in this kinematic region.
Consequently, it is important to have a robust understanding of the sources of uncertainty that affect the DY cross-section over the whole range of invariant masses considered by the experiment. 

The goal of this paper is to provide an assessment of theory uncertainties impacting the inclusive NCDY cross-section in the low-mass region using state-of-the-art theory predictions.
We focus here on uncertainties of QCD origin, and we neglect electroweak corrections, which are expected to essentially factorize from the QCD corrections~\cite{Dittmaier:2014qza,Dittmaier:2015rxo,Dittmaier:2024row}. 
Moreover, we focus our study on the invariant-mass distribution of the produced lepton pair, without imposing any cuts and being inclusive in the additional QCD radiation. While a detailed study should also take the experimental fiducial volume into account, our results will provide valuable information on the theoretical precision that can be reached using state-of-the-art computations. The choice of an inclusive (no lepton selections other than $m_{\ell \bar \ell}$) observable is an idealisation of a more realistic scenario. Both the PDF and scale uncertainties will increase (in a way which we cannot know until computing them).
Applying a similar approach, we additionally quantify how similar effects impact the predictions for the CCDY process.

The structure of the paper is as follows: Section~\ref{sec-theoreticalsetup} introduces the computational framework for both the NCDY and the CCDY processes, detailing the theoretical fixed-order computation framework.  Section~\ref{sec:matching_theory} describes the matching procedure employed to combine massive and massless calculations. In Section \ref{sec:pheno}, we describe the numerical setup and present the differential cross section with respect to the invariant mass, followed by an examination of various sources of uncertainties, starting with those from PDFs.  In section~\ref{sec:conclusions}, we present our conclusions.
Additionally, the paper is accompanied by an appendix which provides further details on the construction of the general mass variable flavour number scheme, with technical details and numerical validations of the adopted procedure.


\section{Theoretical framework for the DY calculation}
\label{sec-theoreticalsetup}

The Drell-Yan (DY) process~\cite{Drell:1970wh} describes the production of a pair of leptonic particles $L_1$ and $L_2$ (that is, charged leptons and/or neutrinos) from the scattering of two protons (P) with momenta $P_1$ and $P_2$ with the center-of-mass energy defined as $S= {(P_1+P_2)}^2$. We denote the invariant mass of the final-state pair by $Q$. The DY process is then given as:
\begin{align}
    {\rm{P}}(P_1) + {\rm{P}} (P_2) \rightarrow L_1 L_2 (Q)+ X\,,
\end{align}
where $X$ denotes the additional QCD radiation.\footnote{More generally this process can be studied in hadron-hadron collisions, here we focus on the case of proton proton (PP) collisions.}  
The \emph{neutral current} (NC) DY process refers to the case where $L_1L_2 = \ell^+ \ell^-$ are charged leptons, while the \emph{charged-current} (CC) DY process will refer to the two cases $L_1L_2 = \ell^- \bar\nu_{\ell}, \ell^+ \nu_{\ell}$.

If the center-of-mass energy of the incoming protons is large enough, the DY process can be described using QCD factorisation. 
The cross-section can then be cast in the form of a convolution of partonic cross-sections, describing the collision of high-energy quarks and gluons, with \emph{parton distribution functions (PDFs)}. More precisely, we can write,
\beq\bsp
Q^2 \, \frac{\rd \sigma^{(n_f,\kappa)}}{\rd Q^2} &\,= \tau \sum_{i,j} \; \int_0^1 \rd x_1\, \rd x_2\,\rd z\,  f_i^{(n_f)} (x_1, \mu_F) \, f_j^{(n_f)} (x_2, \mu_F)\,\\
&\,\qquad\times \eta_{ij}^{(n_f,\kappa)} \big(z, a_s^{(n_f)}(\mu_R)\big) \delta(\tau - x_1 x_2 z)\\
&\,=\tau \sum_{i,j} \mathcal{L}_{ij}^{(n_f)} (\tau, \mu_F) \otimes \eta _{ij}^{(n_f,\kappa)}\big(\tau, a_s^{(n_f)}(\mu_R)\big)\,,
\label{eq:inccrosssection}
\esp\eeq
with $\tau$, $z= \frac{\tau}{x_1 x_2 }$, and $\kappa\in\{\pm1,0\}$, depending on whether the process is mediated by a neutral vector boson ($\kappa=0$) or a $W^{\pm}$ ($\kappa=\pm1$). The sum runs over all (massless) parton species, and we work in a theory of QCD with $n_f$ massless quark flavours.
More specifically, we employ a variable flavour number scheme (VFNS) in which the value of $n_f$ varies~\cite{Collins:1998rz} according to the values of $\mu_{F/R}$.
Here and throughout we use the superscript $n_f$ to specify the heavy-flavour scheme in which the calculation is performed---further details are given in section~\ref{sec:matching_theory}.
The contributions of massive quarks are included through virtual loops or when they are produced in the final-state, but they are not included in the sum over the initial states.\footnote{If one considers an effective theory in which there is an odd number of massive and massless quarks (e.g. after fully decoupling a single massive quark), the axial-current contribution to NCDY is anomalous, and we include appropriate non-decoupling effects, cf.,~e.g.,~\cite{Dicus:1985wx,Ju:2021lah,Chen:2021rft,Duhr:2021vwj}} 
The $f_i^{(n_f)}(x, \mu_F^2)$ and $a_s^{(n_f) }(\mu_R)$ ($a_s= \frac{\alpha_s}{\pi}$)  denote the PDFs and the strong coupling constant renormalised in the $\overline{\textrm{MS}}$ scheme, which depend on the factorization scale $\mu_F$ and the renormalization scale $\mu_R$, respectively. Note that the PDFs and the strong coupling constant and their evolution do depend on the number $n_f$ (and the maximal allowed value $n_{f,\max}$) as dictated by the VFNS.
In the last equality in eq.~\eqref{eq:inccrosssection} we have expressed the cross-section as a convolution with the partonic luminosity,
\begin{align}
\mathcal{L}_{ij}^{(n_f)} (\tau, \mu_F)&= f_i^{(n_f)} (\tau, \mu_F) \otimes f_j^{(n_f)} (\tau,\mu_F)\,,
\end{align}
where we have defined the convolution of two functions $f$ and $g$, 
\beq
(f\otimes g)(\tau) = \int_0^1 f(x_1)\, g (x_2)\,\delta(\tau- x_1 x_2)\, \rd x_1\, \rd x_2 = \int_{\tau}^1 \frac{\rd x}{x}\,f(x)\,g\left(\tfrac{\tau}{x}\right)\,.
\label{eq: etafifj}
\eeq
Finally, the $\eta_{ij}^{(n_f,\kappa)}$ denote the partonic cross-sections for producing a lepton pair from a collision of two partons $i$ and $j$ at LO in the electroweak coupling. 
Note that in the case of NCDY, $\eta_{ij}^{(n_f,0)}$ includes contributions from both $Z$-boson and photon exchange, including interference effects, and the decays of the off-shell vector bosons are described via a Breit-Wigner distribution.
The partonic cross-sections can be expanded into a perturbative series in the strong coupling,
\begin{align}
    \eta _{ij}^{(n_f,\kappa)} \big(z, a_s^{(n_f)}(\mu_R)) = \sum_{k = 0}^\infty a_s^{(n_f)}(\mu_R)^k \,\eta _{ij}^{(n_f,\kappa,k)} (z)\,.
    \label{eq:eta4}
\end{align}

Higher-order QCD corrections to the invariant-mass distribution in eq.~\eqref{eq:inccrosssection} arising from massless quarks have been computed up to N$^3$LO in the strong coupling constant for both NCDY and CCDY processes~\cite{Altarelli:1978id,kubar-andre:1978eri,Altarelli:1979ub,Matsuura:1987wt,Matsuura:1988nd,Matsuura:1988sm,Hamberg:1990np,Matsuura:1990ba,vanNeerven:1991gh,Duhr:2020sdp,Duhr:2020seh,Duhr:2021vwj,Baglio:2022wzu}. Contributions from heavy quarks were computed for the NCDY process in~\cite{Rijken:1995gi,Maltoni:2005wd,FebresCordero:2008ci,Mazzitelli:2024ura} and for the CCDY process in~\cite{Behring:2020uzq,Stelzer:1997ns,Buonocore:2022pqq,Buonocore:2023ljm,Frederix:2011qg}. The N$^3$LO computations have revealed intricate cancellations between contributions from different partonic channels~\cite{Duhr:2020sdp,Duhr:2020seh,Duhr:2021vwj,Baglio:2022wzu} (which are not observed in Higgs production processes at the same order~\cite{Anastasiou:2015vya,Anastasiou:2016cez,Duhr:2020kzd}). In addition, it was observed that the uncertainty from missing higher orders in the perturbative expansion obtained at NNLO by varying the factorization and renormalization scales underestimates the size of the QCD corrections, with a residual scale dependence of a few percent still at N$^3$LO. Based on the available N$^3$LO computations, it is however expected that the perturbative QCD corrections to the DY processes are under control. It is therefore important to address the question what other QCD effects may impact the cross-section at the same level as the QCD scale uncertainty at the percent-level. One of the main goals of our paper is to provide such a study.

It is possible to identify two immediate potential sources of uncertainty that may affect the invariant-mass distribution. 
First, the available N$^3$LO predictions in~\cite{Duhr:2020sdp,Duhr:2020seh,Duhr:2021vwj,Baglio:2022wzu} have been performed using NNLO PDF sets, due to the absence of available N$^3$LO PDFs at the time. In recent years, a lot of progress has been made in determining the four-loop splitting functions governing the evolution of N$^3$LO PDFs~\cite{Basdew-Sharma:2022vya,Falcioni:2023luc,Falcioni:2023vqq,Falcioni:2023tzp,Falcioni:2024xyt,Falcioni:2024qpd,Gehrmann:2023cqm, Gehrmann:2023iah,Kniehl:2025ttz}, and various approximate N$^3$LO (aN$^3$LO) PDF sets have become available~\cite{NNPDF:2024nan,McGowan:2022nag,MSHT:2024tdn}. It is interesting to assess how these new PDF sets impact the theoretical predictions for DY processes.  
Second, at high energies it is typical to treat the heavy-flavour quarks (charm and bottom) as being effectively massless and to perform a resummation of (universal) logarithmically enhanced corrections to the cross-section via an evolution of PDFs and the strong coupling in the VFNS. This approximation neglects power-like corrections in the heavy-flavour quark masses in favour of a resummation of logarithmic corrections. It is not clear how appropriate this approximation is when the description of low invariant mass lepton pairs is desired.
It is therefore mandatory to carefully match the computations for massive and massless quarks in order to obtain reliable predictions. In the remainder of this paper we present a detailed study of aN$^3$LO PDFs and VFNS in DY processes. Before we discuss our phenomenological results, we give a brief review of the VFNS for DY processes.


\section{The massive variable flavor number scheme for the DY process}
\label{sec:matching_theory}

In this section we discuss how predictions for the DY processes can be obtained which include the resummation of heavy-flavour quark mass effects from PDFs and the evolution of $\alpha_s$ (which is standard in the VFNS), but additionally retain power-like corrections in the heavy-flavour quark masses at fixed-order accuracy.
The constructed scheme is often referred to as a general mass or a \textit{massive variable flavour number scheme} (MVFNS).
We follow the approach outlined in~\cite{Gauld:2021zmq}, and in the remainder of this section we present the main features of the method and also provide a numerical validation of its construction.
We focus on the example of the NCDY process, while the CCDY process is detailed in appendix~\ref{appendix:3vs5}.

\subsection{General considerations}

The theoretical framework for the description of the DY process has been introduced in section~\ref{sec-theoreticalsetup}.
This framework allows for the treatment of both massive and massless quark contributions to the cross-section, noting that massive quarks are not included in the sum over initial-state partons in the QCD factorisation formula in eq.~\eqref{eq:inccrosssection} (its inclusion would lead to a non-cancellation of infrared singularities in the initial state, cf.~\cite{Caola:2020xup}).
The VFNS treatment of heavy-flavours, through which the number of massless quark flavours $n_f$ can vary according to $\mu_{F/R}$, allows to account for heavy-flavour quark mass effects in different ways.
For example, in the so-called \emph{4 flavor scheme (4FS)} the maximal value of $n_f$ is $n_{f,\max} = 4$, the bottom quark is always considered massive and only enters as a virtual particle in loops or as a massive final-sate particle. Consequently, there is no bottom quark PDF, and the heavy quark decouples from the running of the strong coupling and the DGLAP evolution. If the mass $m_b$ is much smaller than the invariant mass $Q$, then the cross-section develops large logarithms of the form $\log\tfrac{m_b}{Q}$ order by order in perturbation theory, which may spoil the convergence of the perturbative series.
Instead in the \emph{5 flavor scheme (5FS)}, where $n_{f,\max} = 5$, the bottom quark is treated as massless for $\mu_{F/R} \gtrsim m_b$ and enters the evolution equations for PDFs and the running coupling (through the QCD $\beta$ function). In this case there is a perturbatively generated bottom quark PDF, and the bottom quark contribution is included in the sum over initial-states.
The class of large collinear logarithms $\log\tfrac{m_b}{Q}$ are then resummed to all orders into the (anti-)bottom quark PDF. However, any contributions that are power-suppressed in $m_b$ are neglected (while they are described exactly in the 4FS at fixed-order). 
The above discussion can also be applied to the case of charm quarks to define the \emph{three flavor scheme (3FS)}, where only the $u$, $d$ and $s$ quarks are considered massless, i.e., $n_{f,\max}=3$.

%

The different treatment of heavy-flavours in these schemes results in different contributions being included in the computation of the partonic cross-sections, and the perturbative counting of various channels is also altered accordingly.
To illustrate this for the NCDY process, we show in table ~\ref{fig:diagrams} representative diagrams involving bottom quarks in both the 4FS and 5FS, and their respective perturbative orders in each scheme. The problem of matching 4FS and 5FS predictions for $Z$ boson production in association with heavy-flavour quarks has also been investigated in the context of parton-shower matched predictions~\cite{Bagnaschi:2018dnh,Hoche:2019ncc}. Some other technical developments in this context can also be found in~\cite{Krauss:2017wmx} and comparisons of parton-shower matched predictions in the different schemes can be found, for example, in~\cite{Krauss:2016orf}.  \begin{center}
\begin{table}[!htbp]
\begin{center}
\resizebox{13.25cm}{!}{
\begin{tabular}{|c|c|c|c|c|}
\hline\hline
 & \begin{tikzpicture}[line width=1 pt, scale=0.7]
\draw[fermion] (-2.5,1) -- (0,0);
\draw[fermion](0,0)  -- (-2.5,-1);
\draw[vector] (0,0) -- (2,0);
\node at (-2.7,1) {$b$};
\node at (-2.7,-1) {$\bar{b}$};
\node at (2.5,0.5) {\scriptsize{$\gamma^{*}/Z^{*}\rightarrow l^{+}l^{-}$}};
\end{tikzpicture} & 
\begin{tikzpicture}[line width=1 pt, scale=0.7]
\draw[gluon] (-2.5,1) -- (0,1);
\draw[fermion] (-2.5,-1) -- (0,-1);
\draw[fermion] (0,-1) -- (0,1);
\draw[fermion] (0,1) -- (2,1);
\draw[vector] (2,-1) -- (0,-1);
\node at (-2.7,1) {$g$};
\node at (-2.7,-1) {$b$};
\node at (2.2,1) {$b$};
\node at (2.2,-0.5) {$\gamma/Z\rightarrow l^{+}l^{-}$};
\end{tikzpicture} & 
\begin{tikzpicture}[line width=0.5 pt, scale=0.7]
\draw[gluon] (-2.5,1) -- (0,1);
\draw[gluon] (-2.5,-1) -- (0,-1);
\draw[fermion] (0,-1) -- (0,1);
\draw[vector] (0,0) -- (2,0);
\draw[fermion] (0,1) -- (2,1);
\draw[fermion] (2,-1) -- (0,-1);
\node at (-2.7,1) {$g$};
\node at (-2.7,-1) {$g$};
\node at (2.2,1) {$b$};
\node at (2.2,-1) {$\bar{b}$};
\node at (2.5,0.5) {$\gamma/Z\rightarrow l^{+}l^{-}$};
\end{tikzpicture}
 \begin{tikzpicture}[line width=0.5 pt, scale=0.7]
\draw[fermion] (-2.5,1) -- (0,1);
\draw[fermion] (0,-1) --(-2.5,-1) ;
\draw[fermion] (0,1) -- (0,-1);
\draw[gluon] (0,1) -- (2,1);
\draw[fermion] (2,1) -- (3,1.5);
\draw[fermion] (3,0.5) -- (2,1);
\draw[vector] (2,-1) -- (0,-1);
\node at (-2.7,1) {$q$};
\node at (-2.7,-1) {$\bar{q}$};
\node at (3.2,1.5) {$b$};
\node at (3.2,0.5) {$\bar{b}$};
\node at (2.3,-0.5) {$\gamma/Z\rightarrow l^{+}l^{-}$};
\end{tikzpicture}
& 
\begin{tikzpicture}[line width=0.5 pt, scale=0.7]
\draw[gluon] (-2.5,1) -- (0,1);
\draw[gluon] (-2.5,-1) -- (0,-1);
\draw[gluon] (-1.5,1) -- (-1.5,-1);
\draw[fermion] (0,-1) -- (0,1);
\draw[vector] (0,0) -- (2,0);
\draw[fermion] (0,1) -- (2,1);
\draw[fermion] (2,-1) -- (0,-1);
\node at (-2.7,1) {$g$};
\node at (-2.7,-1) {$g$};
\node at (2.2,1) {$b$};
\node at (2.2,-1) {$\bar{b}$};
\node at (2.5,0.5) {$\gamma/Z\rightarrow l^{+}l^{-}$};
\end{tikzpicture}\\
 \hline &&&&\\
  $\quad$4FS$\quad$ & -- & -- & LO & NLO \\
 $\quad$5FS$\quad$ & LO & NLO & NNLO & N$^3$LO
 \\
  \hline\hline
\end{tabular}}
\caption{\label{fig:diagrams}  Representative diagrams for the NCDY process at different orders in perturbation theory in the 4FS and 5FS.
}
 \end{center}
 \end{table}
 \end{center}

The aim of a MVFNS is to combine both massive and massless approaches (e.g. 4FS and 5FS) such that a combination of resummation and power corrections can be included in the cross-section prediction.
First predictions providing a matching of massive and massless approaches for $Z$-boson production in the bottom-quark fusion ($gg$-induced) channel were performed in~\cite{Forte:2018ovl} using the FONLL scheme~\cite{Forte:2016sja} with 5FS ingredients included up to NNLO (because the corresponding N$^3$LO cross-sections were not yet available at the time). 
A matching of massive and massless schemes generally for heavy-flavour quarks (e.g. charm and bottom quarks) as well as for fully differential kinematics of the reconstructed gauge boson has been given in~\cite{Gauld:2021zmq}. In~\cite{Gauld:2021zmq} the method was applied to the case of NCDY with results valid exactly at $\mathcal{O}(\alpha_s^2)$ and up to $\mathcal{O}(\alpha_s^3)$ in the bottom quark fusion process. 
A goal of this paper is to apply this method in the region of low-invariant dilepton masses in combination with the massless N$^3$LO QCD predictions. A further goal is to apply the method to the case of the CCDY process for the first time.
Before doing so, we provide an outline of the method and show a numerical validation of the procedure.

\subsection{The massive variable flavor number scheme}
\label{sec:powercorrectiondef}

In the following we provide a brief review of the MVFNS of~\cite{Gauld:2021zmq}. We focus on the matching of NCDY process in the 4FS and 5FS to second order in the strong coupling. A discussion of the matching to the 3FS and for the CCDY process can be found in appendix~\ref{appendix:3vs5}. 

The starting point of the matching is the fact that 
the (differential) cross-section in the 4FS and 5FS are the same up to power-suppressed terms in the bottom quark mass. 
We take the following decomposition for the calculation of the differential cross-section in the 4FS:
\begin{align}
    {\rm{d}}\sigma^{(4,0,k)} &= {\rm{d}}\sigma^{(4,0,k)}_{{\rm Light}} + {\rm{d}}\sigma^{(4,0,k)}_M\,,
\end{align}
where ${\rm{d}}\sigma^{(4,0,k)}_{{\rm Light}}$ includes all contributions from light quarks that originate from initial, final, and virtual states. 
In contrast, ${\rm{d}}\sigma^{(4,0,k)}_M$ denotes contributions arising from heavy quarks, encompassing effects in final and/or virtual states.
In the following we focus on the contributions involving bottom quarks\footnote{The contributions of more than one heavy-flavour to the NCDY or CCDY processes do not begin until $\mathcal{O}(\alpha_s^4)$ or $\mathcal{O}(\alpha_s^3)$ respectively, beyond the considered accuracy, which allows to simplify the discussion to the case of a single heavy-flavour at a time.}. 
These contributions can be further decomposed into the following three parts:
\begin{align}
    {\rm{d}}\sigma^{(4,0,k)}_{M,[b]}&={\rm{d}}\sigma_{\ln[m_b]}^{(4,0,k)} + {\rm{d}}\sigma_{n_f,[b]}^{(4,0,k)} + {\rm{d}}\sigma_{pc,[b]}^{(4,0,k)}\,.
    \label{eq:3componentdiffcross4fs}
\end{align}
Here, the subscript $[b]$ acts to denote that we consider those contributions involving the bottom quark.
The first part on the RHS of eq.~\eqref{eq:3componentdiffcross4fs}
${\rm{d}}\sigma_{\ln[m_b]}^{(4,0,k)}$ captures the logarithmically enhanced terms in the limit $m_b \to 0$; the second ${\rm{d}}\sigma_{n_f,[b]}^{(4,0,k)}$ accounts for the constant terms in this limit; and the third ${\rm{d}}\sigma_{pc,[b]}^{(4,0,k)}$ represents the power-suppressed corrections as $m_b \to 0$.
The power correction terms ${\rm{d}}\sigma_{pc,[b]}^{(4,0,k)}$ are absent in the 5FS differential cross-section, and the isolation of these corrections enables the matching of the predictions in the 4FS and 5FS through a simple additive procedure.
This isolation is achieved by the independent computation of the quantity in the left-hand side and the first two terms in the right-hand side of eq.~\eqref{eq:3componentdiffcross4fs}, and re-arranging for ${\rm{d}}\sigma_{pc,[b]}^{(4,0,k)}$.
For the eventual combination of the schemes, it is also convenient to evaluate all contributions with 5FS inputs.
This can be achieved by re-expressing each of the terms appearing in eq.~\eqref{eq:3componentdiffcross4fs} in terms of 5FS inputs as
\begin{align} \nonumber
    {\rm{d}}\sigma^{(4,0,k)}_{i,[b]}&=\left( {\rm{d}}\sigma^{(5,0,k)}_{i,[b]} + {\rm{d}}\sigma^{(5,0,k)}_{i,[4\leftrightarrow5]}\right)+\mathcal{O}(\alpha_s)\,,\\
    &={\rm{d}}\sigma^{(5,0,k)}_{i,[b]}+\mathcal{O}(\alpha_s)\,,
    \label{eq:3componentdiffcross}
\end{align}
with the first subscript $i=M,\ln[m],n_f,pc$ denoting the component of the heavy-flavour cross-section appearing in eq.~\eqref{eq:3componentdiffcross4fs}, and the additional subscript $[b]$ specifying the heavy flavour being considered (the bottom quark in this case).
%
The compensation terms ${\rm{d}}\sigma^{(5,0,k)}_{i,[4\leftrightarrow5]}$ correct for any inconsistencies introduced with the use of 5FS inputs, order-by-order in the strong coupling.
To the (relative) first order, the required corrections terms impact the gluon PDF and the strong coupling---see eqs. (3.15,3.16) of~\cite{Cacciari:1998it}. 
In the following, we outline how the terms appearing in eq.~\eqref{eq:3componentdiffcross} are evaluated.


\paragraph{The massive contributions, ${\rm{d}}\sigma^{(5,0,k)}_{M}$.}
The massive quark contributions to the NCDY process begin at $\mathcal{O}(\alpha_s^2)$. Those include contributions in which the heavy-flavour quarks are produced in the final state (see table~\ref{fig:diagrams}), but also contributions from lower-multiplicity subprocesses through double-virtual corrections and ultraviolet (UV) renormalization counterterms. The two-loop corrections that contribute to $\mathcal{O}(\alpha_s^2)$ have been recently computed in~\cite{Behring:2020uzq} (see also~\cite{Kniehl:1989kz}).
The inclusion of all of these heavy-flavour contributions at a given fixed-order, and channel by channel, is critical for the procedure.

At this point, we note that while all heavy-flavour contributions to the NC process are known at $\mathcal{O}(\alpha_s^2)$, several ingredients are missing at $\mathcal{O}(\alpha_s^3)$.
For example, required two-loop three-parton and three-loop two-parton amplitudes involving internal massive quark lines relevant for the $q\bar q$ and $q g$ induced sub-processes are currently unknown.
While all ingredients for the $gg$ channel are known at $\mathcal{O}(\alpha_s^3)$, we have chosen not to include these partial results (see also the discussion towards the end of this paragraph).

In this work the massive $\mathcal{O}(\alpha_s^2)$ contributions are computed with a Monte Carlo generator developed in~\cite{Gauld:2021zmq}, which relies on external libraries from OpenLoops~\cite{Buccioni:2019sur} and the Vegas algorithm implemented in the Cuba library~\cite{Hahn:2004fe}.
An exception being a subset of massive corrections to the CC process at $\mathcal{O}(\alpha_s^2)$ which are instead taken from {\tt{MCFM-10.3}}~\cite{Campbell:2015qma,Campbell:2019dru}.


\paragraph{The logarithmic cross-section.} For QCD inclusive and/or IRC safe observables in hadron-hadron collisions, the cross-section prediction contains a logarithmic sensitivity to the heavy-flavour quark mass of a collinear origin. At fixed-order accuracy, these logarithmic corrections are described by the heavy-quark decoupling relations for both the PDFs and the strong coupling $\alpha_s$.
Decoupling relations are used to describe parameters such as $\alpha_s$ and PDFs in a theory with a massive quark (4FS) relative to an effective theory where the corresponding quark is treated as massless (5FS) at a fixed order.
For example, they describe the transition of these quantities across heavy-flavour thresholds in the VFNS.
These decoupling relations, in combination with massless partonic cross-sections of lower perturbative order, can therefore be used to describe the logarithmic behaviour of the massive calculation at fixed-order accuracy.
In our calculation we choose to express these relations in terms of PDFs and the strong coupling in the 5FS scheme. 

%
%
Accordingly, the decoupling relation for the strong coupling constant at the bottom quark threshold (i.e. the transition between $n_f=4$ and $n_f=5$), matched at a scale $\mu$, is given by~\cite{Chetyrkin:1997sg}:
\beq\bsp
\label{oms}
\frac{a_s^{(4)}(\mu)}{a_s^{(5)}(\mu)} &= 1 - a_s^{(5)}(\mu) \frac{L_Q}{6} 
+ a_s^{(5)}(\mu)^2\left(\frac{L_Q^2}{36} - \frac{19}{24}L_Q -\frac{7}{24}\right) \\
&\quad + a_s^{(5)}(\mu)^3\left(-\frac{L_Q^3}{216} - \frac{131}{576}L_Q^2 - \frac{6885}{1728}L_Q + C_3\right) + \mathcal{O}\big(a_s^{(5)}(\mu)^4\big)\,,
\esp\eeq
where ${L_Q} = \ln \frac{\mu^2}{m_b^2}$, and we defined,
\beq\bsp
C_3 &= -\frac{80507}{27648}\zeta(3) 
-\frac{2}{3}\zeta(2)\left(\frac{1}{3}\ln2 + 1\right) 
-\frac{58933}{124416} 
+\frac{4}{9}\left(\zeta(2) + \frac{2479}{3456}\right)\,.
\esp\eeq
Decoupling relations for the PDFs are instead provided in terms of operator matrix elements (OMEs). They describe transitions between partonic states $i \rightarrow j$ and are differential in a collinear variable.
For example, PDFs with $n_f=5$ can be expressed in terms of those with $n_f=4$ according to
\begin{align}
f_i^{(5)} = \sum_{j=-4}^4 K_{ij}\big(L_Q, a_s^{(4)}\big) \otimes f_j^{(4)}, \quad -5 \leq i \leq 5\,,
\end{align}
where we suppress the explicit dependence on $\mu_F$. The fixed-order expansion of the bottom-quark PDF in terms of 5FS inputs is then expressed as:
\begin{align}
f_b^{(5)} = f_{\bar{b}}^{(5)} = a_s^{(5)}\, A_{bg}^{(1)} \otimes f_g^{(5)} 
+ a_s^{(5)2} \bigg[A_{bg}^{(2)} \otimes f_g^{(5)} + \sum_{\substack{i=-4\\ i \neq 0}}^4 A_{bi}^{(2)} \otimes f_i^{(5)}\bigg] + \mathcal{O}(a_s^{(5)3})\,,
\label{eq:bpdf5}
\end{align}
where $A_{ij}^{(k)}$ are OMEs at perturbative order $a_s^{(5)k}$, which up to $k=2$ can be found in~\cite{Buza:1996wv,Bierenbaum:2009zt}. The OMEs are also known for $k=3$ and they can be found in~\cite{Blumlein:2021lmf,Blumlein:2021xlc,Ablinger:2019gpu,Allwicher:2022gkm}.
Substituting the decoupling relations,
we obtain the following expression for ${\rm{d}}\sigma_{\ln[m_b]}^{(5,0,k)}$ up to second order in the strong coupling:
\begin{align}
{\rm{d}}\sigma_{\ln[m_b]}^{(5,0,k)} = \tau \sum_{i,j=-4}^4 \mathcal{L}_{ij}^{(5)}(\tau, \mu_F) \otimes \delta \eta_{ij,\ln[m_b]}^{(5,0,k)}\big(\tau, \mu_F, \mu_R, L_Q, \alpha_s^{(5)}, m_b\big)\,.\label{eq:dsigma_logm}
\end{align}
For the $gg$ induced channel this leads to the expressions,
\beq\bsp
\delta \eta_{gg,\ln[m_b]}^{(5,0,2)} &= 2 A_{bg}^{(1)} \otimes A_{bg}^{(1)} \otimes \eta_{q\bar{q}}^{(5,0,0)} + 4 A_{bg}^{(1)} \otimes \eta_{qg}^{(5,0,1)}\,.
\label{eq:powerexpansionofAs}
\esp\eeq
On the other hand, the $q\bar{q}$-induced channel yields the following logarithmic corrections:
\beq\bsp
\delta \eta_{q\bar q,\ln[m_b]}^{(5,0,2)} &= \left(A_{qq}^{(2)} + A_{\bar q\bar q}^{(2)}\right) \otimes \eta_{q\bar{q}}^{(5,0,0)} +  a_s^{(5)}(\mu) \frac{L_Q}{6} \eta_{q\bar{q}}^{(5,0,1)}\,,
\label{eq:siglogqqb}
\esp\eeq
It is to be noted that the logarithmic corrections in eq.~\eqref{eq:powerexpansionofAs} were derived using only the PDF decoupling relations given in eq.~\eqref{eq:bpdf5}, whereas those in eq.~\eqref{eq:siglogqqb} additionally include a correction arising from the strong coupling decoupling relations given in eq.~\eqref{oms}. Furthermore, the OMEs given in eq.~\eqref{eq:siglogqqb} correspond to the non-singlet contributions arising from light quark interactions, as described in eq (3.6) of \cite{Buza:1996wv}.

All the functions entering the convolutions in the right-hand side of eq.~\eqref{eq:powerexpansionofAs} can be expressed in terms of harmonic polylogarithms~\cite{Remiddi:1999ew}, and we have performed all convolutions analytically using \textsc{PolyLogTools}~\cite{Bauer:2000cp, Duhr:2019tlz}. 
%

\paragraph{The massless constant.}
The mass independent constant ${\rm{d}}\sigma_{n_f}^{(5,0,k)}$ can be directly extracted from the massless partonic cross-sections.
It corresponds to the component of the partonic cross-section arising from a single heavy-flavour quark---i.e. the massless counterpart of the previously described ``massive contributions". 
These contributions can be extracted from the fully massless calculation which is available up to N$^3$LO ($k=3$) in the literature~\cite{Altarelli:1978id,kubar-andre:1978eri,Altarelli:1979ub,Matsuura:1987wt,Matsuura:1988nd,Matsuura:1988sm,Hamberg:1990np,Matsuura:1990ba,vanNeerven:1991gh,Duhr:2020sdp,Duhr:2020seh,Duhr:2021vwj}. We evaluate them directly using the publicly available code {\tt{n3loxs}}~\cite{Baglio:2022wzu}.

\paragraph{The power corrections.}
In the above, it has been detailed how to separately construct the massive, logarithmic, and massless constant cross-sections contributions to differential quantities.
With knowledge of these cross-sections, according to Eq.~\eqref{eq:3componentdiffcross4fs} and Eq.~\eqref{eq:3componentdiffcross}, the contribution arising from power corrections in the heavy quark mass can be extracted/isolated as
\begin{align}
{\rm{d}}\sigma_{pc,[b]}^{(5,0,k)} &=
{\rm{d}}\sigma^{(5,0,k)}_{M,[b]} - \left( {\rm{d}}\sigma_{\ln[m_b]}^{(5,0,k)} + {\rm{d}}\sigma_{n_f,[b]}^{(5,0,k)} \right)\,,
    \label{eq:pc}
\end{align}
where the term in parentheses corresponds to the massless limit of the massive cross-section.
In the limit of $m_b\to$ the RHS (and hence the LHS) of Eq.~\eqref{eq:pc} vanishes.

\paragraph{The combination of massive and massless cross-sections.}
The cross-section prediction in the MVFNS, which combines massive and massless approaches, can be defined with a simple additive procedure according to
\begin{align}
{\rm{d}}\sigma^{({\rm MVFNS},0)} = {\rm{d}}\sigma^{(5,0)} +  {\rm{d}}\sigma_{pc,[b]}^{(5,0)} \,.
    \label{eq:MVFNS}
\end{align}
In this way, the matched prediction improves upon the 5FS prediction, denoted by ${\rm{d}}\sigma^{(5,\kappa)}$, by including the power suppressed terms in ${\rm{d}}\sigma_{pc,[b]}^{(5,\kappa)}$.
In the limit of $m_b\to0$ the power suppressed terms in Eq.~\eqref{eq:MVFNS} vanish and the 5fs prediction is recovered.
While the above discussion focusses on the inclusion of $b$-quark mass corrections for the NCDY process, the inclusion of charm-quark mass effects can be performed in a similar way.
The same formalism also applies to the CCDY case.
Note also that as the power corrections are computed independently of the 5FS prediction, meaning that the perturbative orders of the two contributions on the RHS of Eq.~\eqref{eq:MVFNS} need not be the same.

The construction of the MVFNS given in this section can be qualitatively thought of as the addition of power-correction terms to the fully massless prediction.
Given that all corrections that enter the matching procedure are evaluated with the same set of 5FS inputs, it is also consistent to view the matching procedure from another view point: the massive calculation forms the baseline and the resummed corrections are added to it (with the overlap removed from the massless prediction).

\paragraph{Differential quantities and the extraction of power corrections.}
In the following we will typically discuss phenomenological results for the binned invariant-mass distribution, obtained by integrating eq.~\eqref{eq:inccrosssection} between two scales $Q_{\textrm{min}}$ and $Q_{\textrm{max}}$,
\beq
\Sigma^{(n_f,\kappa)}(Q_{\textrm{min}},Q_{\textrm{max}}) = \int_{Q_{\textrm{min}}^2}^{Q_{\textrm{max}}^2}\rd Q^2\,\frac{\rd \sigma^{(n_f,\kappa)}}{\rd Q^2}\,,
\eeq
where we suppress the dependence of all quantities on the renormalization and factorization scales. By expanding the integrand in perturbation theory, we obtain the perturbative expansion of the bins,
\beq\bsp
\Sigma^{(n_f,\kappa)}&(Q_{\textrm{min}},Q_{\textrm{max}}) = \Sigma^{(n_f,\kappa,0)}(Q_{\textrm{min}},Q_{\textrm{max}}) + a_s^{(n_f)}(\mu_R)\,\Sigma^{(n_f,\kappa,1)}(Q_{\textrm{min}},Q_{\textrm{max}})\\
&\,+ a_s^{(n_f)}(\mu_R)^2\,\Sigma^{(n_f,\kappa,2)}(Q_{\textrm{min}},Q_{\textrm{max}})+\mathcal{O}\big(a_s^{(n_f)}(\mu_R)^3\big)\,,
\esp\eeq
and we define
\beq
\Sigma^{(n_f,\kappa)}_{\textrm{N$^k$LO}}(Q_{\textrm{min}},Q_{\textrm{max}}) = \sum_{l=0}^k a_s^{(n_f)}(\mu_R)^l\,\Sigma^{(n_f,\kappa,l)}(Q_{\textrm{min}},Q_{\textrm{max}})\,.
\eeq

\begin{figure}[!htbp]
     \centering
\includegraphics[width=1\linewidth]{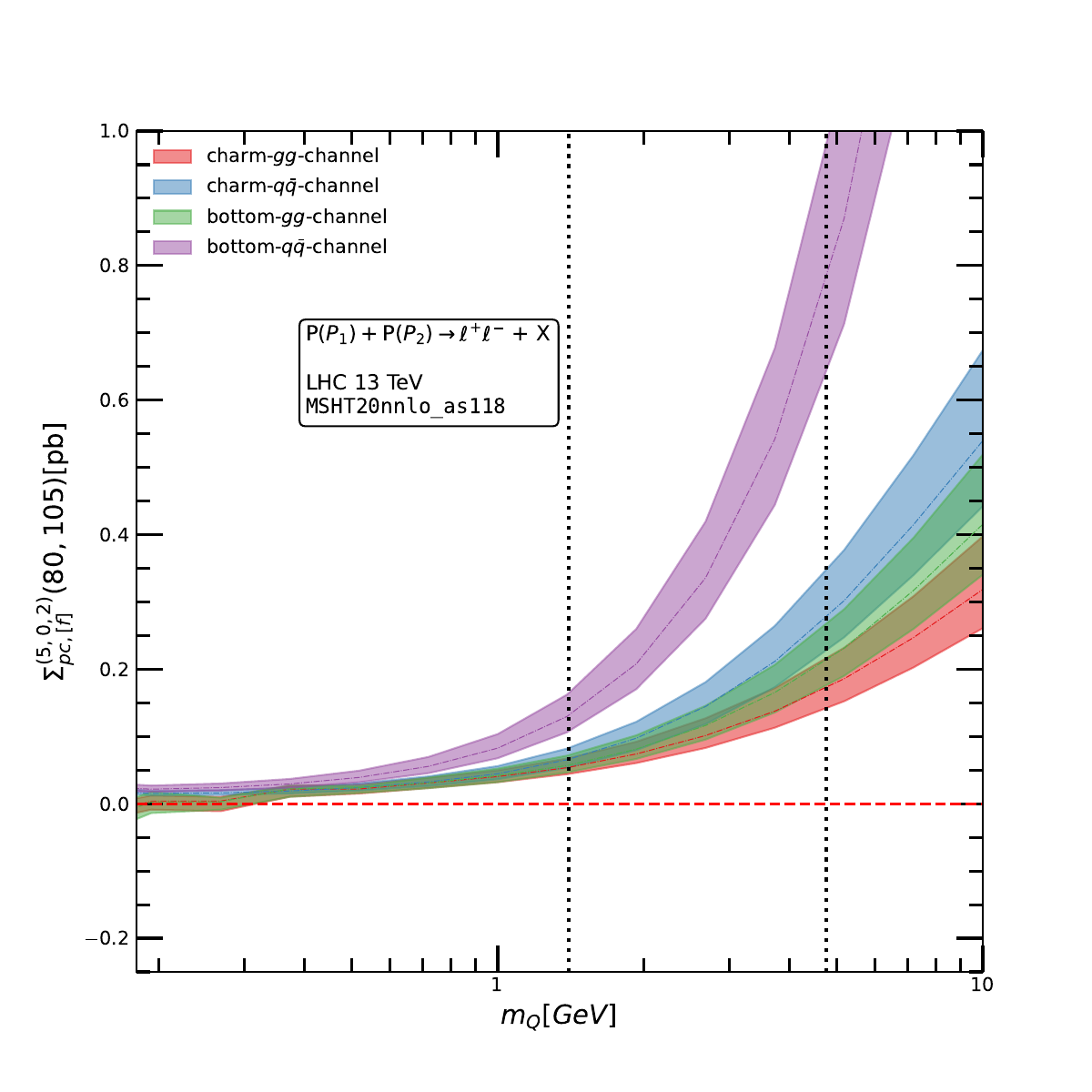}
    \caption{Power-corrections to the NCDY process up to NNLO accuracy defined in eq.~\eqref{eq:Sigma_power_corrections} for the $Q$-bin (80 GeV, 105 GeV). The bands represent represent the 7-point variation of $\mu_R$ and $\mu_F$ around the central (dynamic) scale $\mu_R=\mu_F=Q$ (see section~\ref{sec:th_uncertainties}). We observe that the power corrections vanish in the limit $m_Q \rightarrow 0$. 
     \label{fig:enter-label3}}
 \end{figure}

Using the previous results, we can extract the power-suppressed terms ${\rm{d}}\sigma_{pc}^{(5,0,k)}$ that contribute to the invariant-mass distribution. As a validation of our computation, we show in fig.~\ref{fig:enter-label3} the binned cross-section
\begin{align}\label{eq:Sigma_power_corrections}\nonumber
\Sigma_{{i},[f]}^{(5,0,2)}(Q_{\textrm{min}},Q_{\textrm{max}}) = \int_{Q_{\textrm{min}}^2}^{Q_{\textrm{max}}^2}\rd Q^2\,\frac{\rd \sigma^{(5,0,2)}_{i,[f]}}{\rd Q^2}\,,\qquad i &= M, \,\ln[m],\, n_f,\, pc\,,\\
   f &= c,b\,,
\end{align}
as a function of $m_Q$.  
%

We now show separately the extracted power corrections for charm and bottom quarks, and additionally show the corrections in the $gg$ and $q\bar q$ induced channels (we use the same numerical setup as for our phenomenological study in section~\ref{sec:pheno}).
Importantly, we observe that (within numerical uncertainties) the power corrections vanish in the limit as $m_Q\to0$. 
A few notable comments are worth making regarding these tests:
\begin{itemize}
    \item They are preformed by evaluating the massive and logarithmic cross-sections for decreasing values for $m_Q$ (at discrete values of $m_Q$). In this limit, the corrections of logarithmic origin become enhanced, while any power corrections proportional to the quark mass as part of the massive calculation vanish.  
    \item In this limit, a large cancellation occurs between the full massive computation and the direct logarithmic calculation. It is therefore important to achieve relatively high numerical accuracy for the predictions in this limit. This is typically most challenging for the massive calculation which, for some contributions, involves the integration over the phase-space of final-state quarks produced in pseudo-collinear limits. For example, when $m_Q \sim 0.2$ GeV, the relative precision of the massive calculation is at the level of a few $\sim$\textperthousand. 
    \item Finally, the power corrections in the light-quark induced channel exceed those of the $gg$ channel by several factors (despite receiving a lower PDF luminosity).  
\end{itemize}

\begin{figure}
\centering
\begin{minipage}[t]{1\textwidth}
  \centering  \includegraphics[width=1\linewidth]{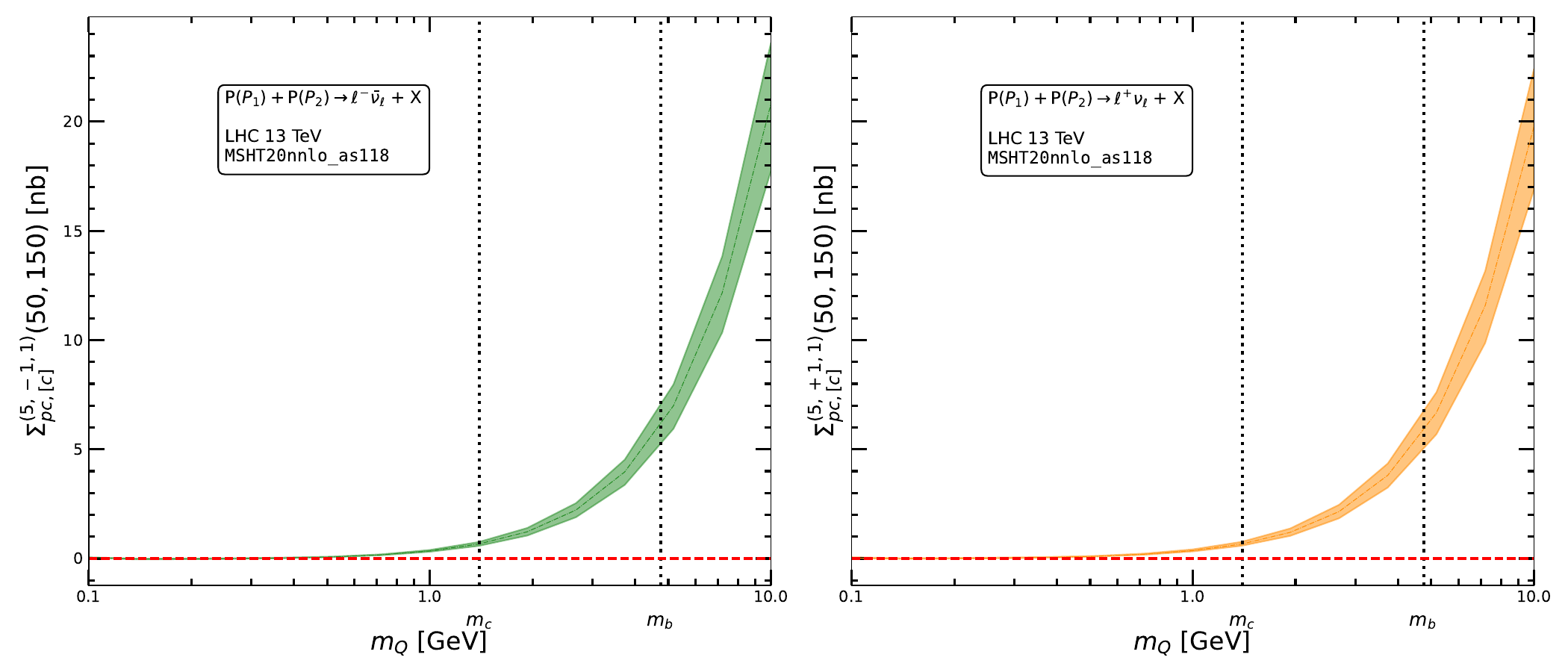}
\end{minipage}
\caption{Power-corrections to the CCDY process up to NLO accuracy for the $Q$-bin (50 GeV, 150 GeV). The bands represent the 7-point variation of $\mu_R$ and $\mu_F$ around the central (dynamic) scale $\mu_R=\mu_F=Q$ (see section~\ref{sec:th_uncertainties}). We observe that the power corrections vanish in the limit $m_Q \rightarrow 0$.}
\label{fig:CDY_LO_pc}
\end{figure}

A similar validation for the CC process at LO is provided in fig.~\ref{fig:CDY_LO_pc}. 
More details on the formula required to construct the MVFNS for both the NC and CC processes are summarised in appendix~\ref{appendix:3vs5}. 
This includes results for the CC process up to $\mathcal{O}(\alpha_s^2)$.

\section{Phenomenology}
\label{sec:pheno}

In this section, we present the main results of our paper, namely a phenomenological study of various sources of QCD uncertainties that may affect the DY invariant-mass distribution. We start by discussing our numerical setup, and then we discuss uncertainties related to the choice of PDFs and the treatment and value of quark masses in turn.

\subsection{Numerical setup}
Unless stated otherwise, in our analysis we work with the {\tt{MSHT20nnlo\_as118}} PDF sets.
As default, we use the charm and bottom quark masses provided together with the corresponding PDF fits, and we make the following choice for the top quark: 
\begin{align}\label{eq:quarkmassvalues}
    m_c = 1.4~\text{GeV}, \quad m_b = 4.75~\text{GeV}, \quad m_t = 173.0~\text{GeV}\,.
\end{align}
We also use the $\alpha_s$ grid corresponding to $\alpha_s(m_Z) = 0.118$ provided alongside the PDF sets.
For the electroweak (EW) scheme, we use the Fermi constant $G_{F}^{\mu}$, the $W$ boson mass $M_W$, and the $Z$ boson mass $M_Z$ as inputs to derive $s_W^2$, $c_W^2$, and $\alpha$. The on-shell values for these parameters (as well as the widths of the bosons) are taken from the PDG~\cite{10.1093/ptep/ptaa104}:
\begin{align}
    m_Z^{\text{os}} &= 91.1876~\text{GeV}\,, \quad \Gamma_Z^{\text{os}} = 2.4952~\text{GeV}\,, \quad
    m_W^{\text{os}} = 80.377~\text{GeV}\,, \quad \Gamma_W^{\text{os}} = 2.085~\text{GeV}\,, \nonumber\\
    G_F^{\mu} &= 1.1663787 \cdot 10^{-5}~\text{GeV}^{-2}\,.
\end{align}
To compute the pole values (used in propagators, etc.), we use the conversion formula from eq. (550) of~\cite{Denner:2019vbn}:
\begin{align}
    m^{\text{pole}} &= \frac{m^{\text{os}}}{\sqrt{1 + \left( \frac{\Gamma^{\text{os}}}{m^{\text{os}}} \right)^2}}\,, \quad 
    \Gamma^{\text{pole}} = \frac{\Gamma^{\text{os}}}{\sqrt{1 + \left( \frac{\Gamma^{\text{os}}}{m^{\text{os}}} \right)^2}}\,.
\end{align}
Using these, the values for the weak mixing angle are derived from eq. (565) of~\cite{Denner:2019vbn},
\begin{align}
    \sin^2\theta_W =  1 - \frac{\mu_W^2}{\mu_Z^2}\,,
\end{align}
where $\mu_W$ and $\mu_Z$ are the (complex) $W$ and $Z$ mass parameters in the complex mass scheme.
Finally, following eq. (3.27) of~\cite{Buccioni:2019sur}, the value of $\alpha$ is computed as:
\begin{align}
    \alpha|_{G_{\mu}} = \frac{\sqrt{2}}{\pi} G_F^{\mu} |\mu_W^2\, \sin^2\theta_W|\,.
\end{align}
%
%
For the CCDY process, we fix the magnitude of the CKM matrix elements with the following simplifications:
\begin{align}
    |V_{ud}| &= 0.974460\,, & |V_{us}| &= 0.224561\,, \nonumber \\
    |V_{cd}| &= 0.224561\,, & |V_{cs}| &= 0.974460\,, \nonumber \\
    |V_{tb}| &= 1.0\,,
\end{align}
with all other CKM matrix elements set to zero.

\begin{figure}
    \centering
\includegraphics[width=1.0\linewidth]{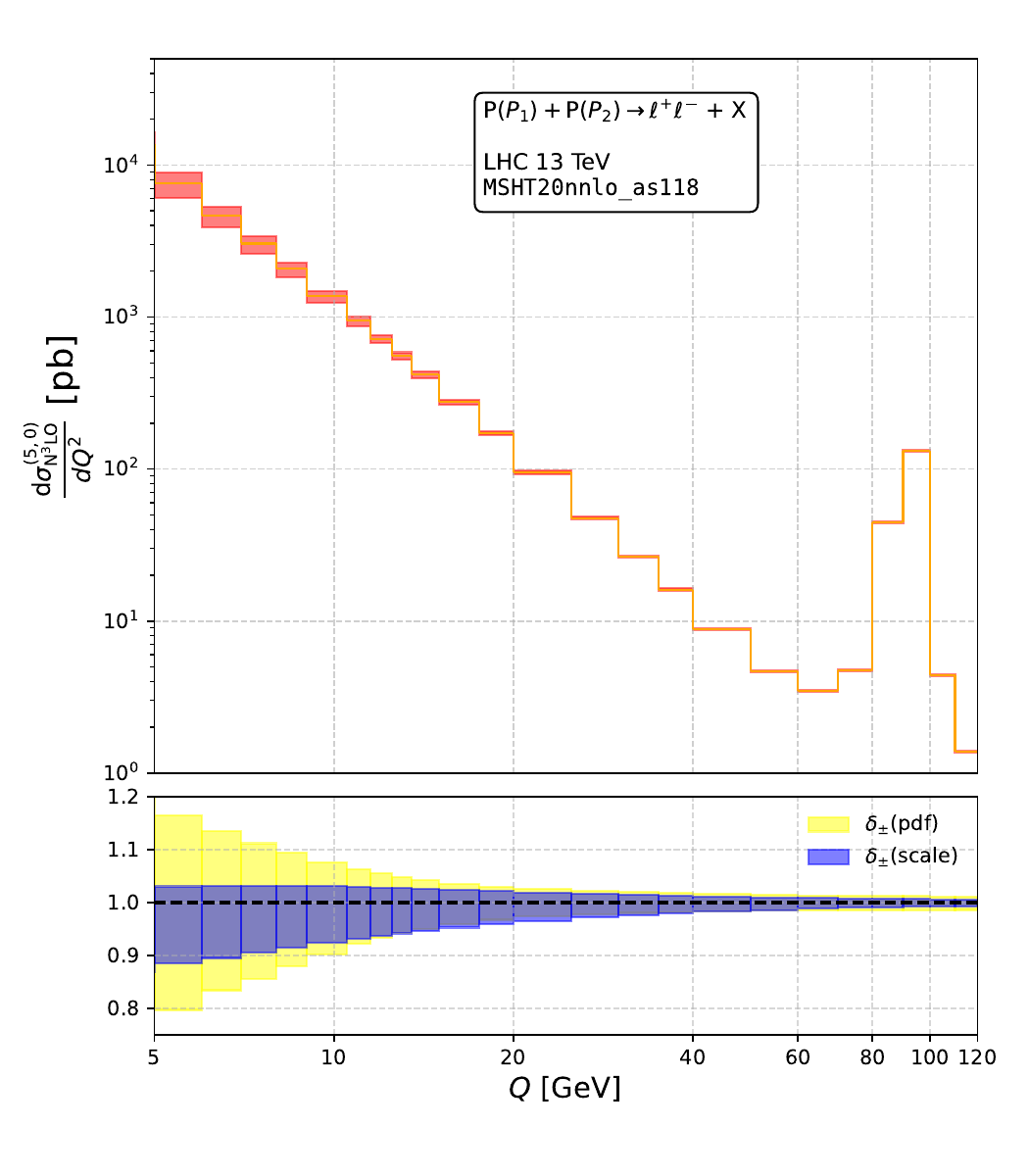}
    \caption{The upper panel shows the scale variation of the NCDY cross-section at $\text{N}^3$LO using {\tt{MSHT20nnlo\_as118}} PDF sets. In the lower panel we plot the variation of the scale variation {\textit{w.r.t}} to the central scale $\mu_R=\mu_F=Q$ and the PDF variation {\textit{w.r.t}} to the central PDF set for $\mu_R=\mu_F=Q$.}
    \label{fig:plot1}
\end{figure}


\subsection{Theory uncertainties}
\label{sec:th_uncertainties}

We start by discussing the uncertainties on the DY invariant-mass distribution in the 5FS which arise due to the truncation of the perturbative series and the uncertainties associated with the PDFs. While the cross-section is independent of the choice of the renormalization and factorization scales to all orders in perturbation theory, the truncation of the perturbative series induces a residual dependence on these scales order by order. Since the PDFs also depend on the factorization scale, the study of the scale dependence is intimately related to the uncertainty coming from the choice and the fitting procedure for the PDFs, and we therefore discuss them simultaneously. We note that the dependence on the perturbative scales is often used as an estimator for the missing higher orders due to the truncation of the perturbative series. 
Approaches beyond scale variation also exist, and have been discussed in~\cite{Cacciari:2011ze,Bagnaschi:2014wea,Bonvini:2020xeo,Duhr:2021mfd,Tackmann:2024kci}. 

A detailed study of the scale dependence and PDF uncertainties for the invariant-mass distribution in DY processes up to N$^3$LO was already performed in~\cite{Duhr:2020sdp,Duhr:2020seh,Duhr:2021vwj,Baglio:2022wzu}. Those studies, however, were based on NNLO PDF sets, due to the absence of N$^3$LO sets. Recently, approximate N$^3$LO (aN$^3$LO) PDFs have been released~\cite{McGowan:2022nag,NNPDF:2024nan,MSHT:2024tdn}, and we therefore repeat the analysis of the scale and PDF uncertainties using these state-of-the-art sets. Unless stated otherwise, we use the {\tt {MSHT20an3lo\_as118}} PDF sets from~\cite{McGowan:2022nag}. We note that there are two aN$^3$LO sets provided by MSHT20.
MSHT20 Set I for a$\text{N}^3$LO ($H_{ij}$ + $K_{ij}$) is referred to as {\tt {MSHT20an3lo\_as118}}-I PDF and a$\text{N}^3$LO  PDF MSHT20 Set II for a$\text{N}^3$LO  ($H'_{ij}$ ) as {\tt {MSHT20an3lo\_as118}}-II PDF. The difference between the two sets arises from how correlations are treated between the $K$-factor parameters for each process and the other PDF and theoretical parameters. Although these correlations are generally small—with a few exceptions—the key distinction lies in whether the `pure' theoretical parameters (such as splitting functions and transition matrix elements) are included within the standard MSHT20 eigenvector analysis, with the $K$-factor parameters decorrelated. This treatment leads to the formation of two separate sets. Further details can be found in section 8 of~\cite{McGowan:2022nag}.

\subsection{Scale uncertainty}
Let us start by discussing the scale dependence. We vary  the perturbative scales $\mu_F$ and $\mu_R$  by a factor of two around the central scales $(\mu_F^0,\mu_R^0)$ while respecting the constraint: \begin{equation}
        \frac{1}{2}\leq \frac{\mu_R}{\mu_F}\leq 2\,.
    \end{equation}
More precisely, we consider the \emph{7-point scale variation}, which allows us to assign the  scale uncertainty
\beq\bsp
\delta_+(\textrm{scale}) &=   \max_{(k_1,k_2)\in S_7} \Big[ \Sigma_{\textrm{N$^3$LO}}^{(5,\kappa)}(k_1\mu_R^0,k_2\mu_F^0)\Big] - \Sigma_{\textrm{N$^3$LO}}^{(5,\kappa)}(\mu_F^0,\mu_R^0)\,  , \\
\delta_-(\textrm{scale}) &=   \min_{(k_1,k_2)\in S_7} \Big[ \Sigma_{\textrm{N$^3$LO}}^{(5,\kappa)}(k_1\mu_R^0,k_2\mu_F^0)\Big]  - \Sigma_{\textrm{N$^3$LO}}^{(5,\kappa)}(\mu_F^0,\mu_R^0)\,  , 
\label{eq:scale_def}
\esp\eeq
where for brevity we suppress the dependence on the bin range $(Q_{\textrm{min}},Q_{\textrm{max}})$ and we defined
\beq
S_7 = \big\{(\tfrac{1}{2},\tfrac{1}{2}), (\tfrac{1}{2},1), (1,\tfrac{1}{2}),(1,1),(2,1),(1,2),(2,2)\big\}\,.
\eeq
We can of course define the scale uncertainty $\delta_\pm(\textrm{scale})$ also directly for the invariant-mass distribution in eq.~\eqref{eq:inccrosssection} by substituting $Q^2\tfrac{\rd \sigma^{(5,\kappa)}}{\rd Q^2}$ in eq.~\eqref{eq:scale_def}. 

\begin{figure}
    \centering
\includegraphics[width=0.7\linewidth]{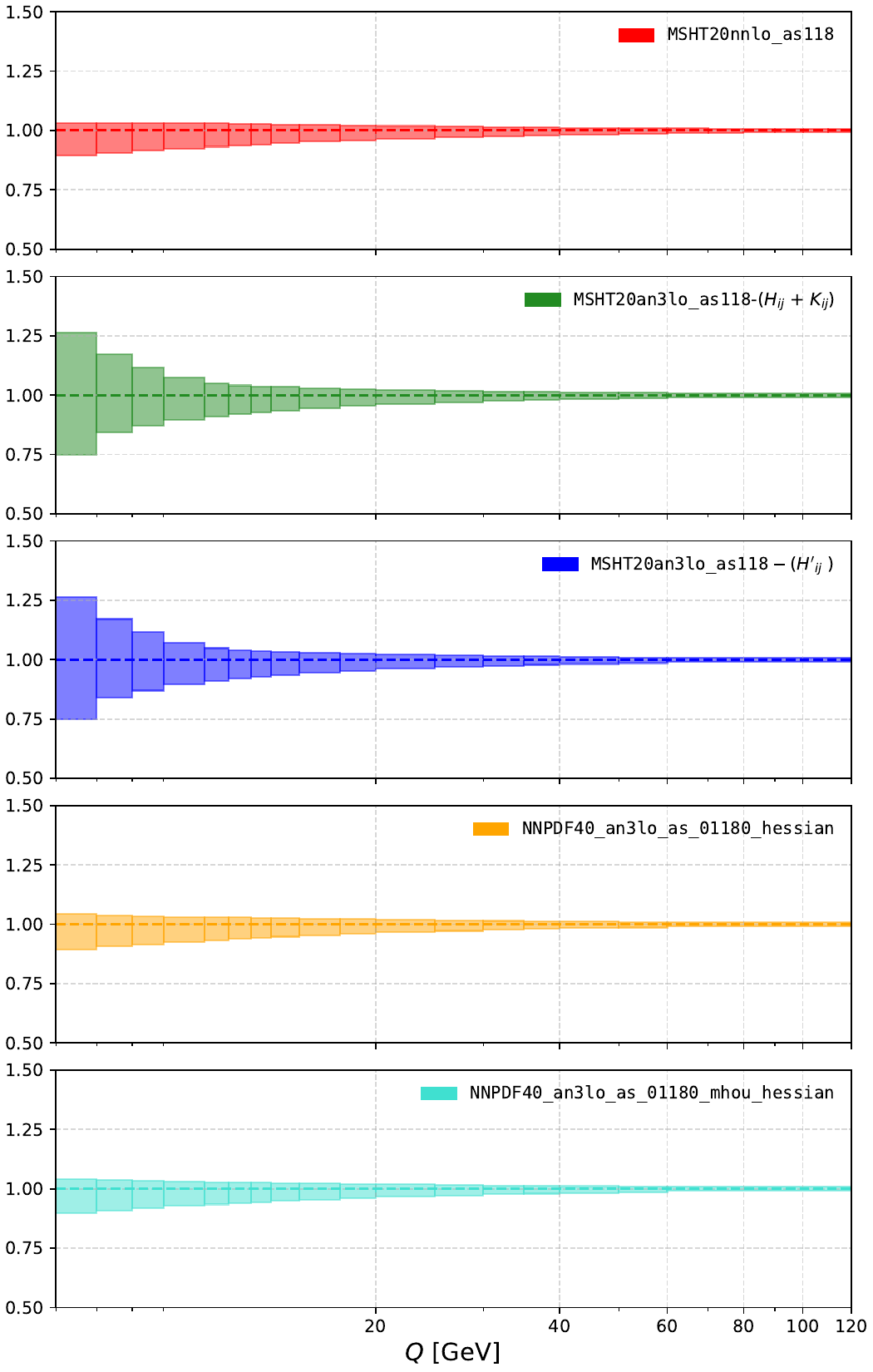}
    \caption{The scale dependence of the NCDY process at $\text{N}^3$LO in the 5FS for  different PDF sets for the $Q$ in the range (4 GeV, 120 GeV).  The uncertainty band was obtained from a 7-point variation around the central scale $\mu_F^0=\mu_R^0=Q$, where $Q$ is integrated between $Q_{\textrm{min}}$ and $Q_{\textrm{max}}$. The $y$-axis corresponds to the ratio  of $\delta_\pm(\text{scale})$ \emph{w.r.t.} the central scale.}
    \label{fig:newscale1}
\end{figure}

In fig.~\ref{fig:newscale1} we show the scale uncertainties for the {\tt {MSHT20nnlo\_as118}} and aN$^{3}$LO PDF sets as a function of the invariant mass $Q\in (4\textrm{ GeV}, 120\textrm{ GeV})$. The ratios are shown for each of the 7-point scale variations, computed relative to the central scale choice of $\mu_R = \mu_F = Q$ for the corresponding central PDF set.    
We observe that at low values of $Q$, the scale uncertainty associated with the aN$^{3}$LO PDF set from MSHT20 is larger than that of the NNLO PDF sets. To determine whether this increased uncertainty is a general feature of N$^{3}$LO PDFs or specific to the MSHT20 analysis, we also include results from the aN$^{3}$LO NNPDF sets in fig.~\ref{fig:newscale1}. In NNPDF also there are two different PDF sets considered: the {\tt NNPDF40\_an3lo\_as\_01180\_hessian} and  {\tt NNPDF40\_an3lo\_as\_01180\_mhou\_hessian} sets.  The latter incorporates uncertainties associated with missing higher-order corrections (MHOU), which are estimated via scale variation and implemented through a theory covariance matrix formalism~\cite{NNPDF:2024nan}. We find that in the low-$Q$ region, the scale variation bands for the NNPDF sets are significantly narrower compared to those of the MSHT20 aN$^3$LO set.

 In tables~\ref{tab: scale_uncertainty_NNLO},~\ref{tab: scale_uncertainty_aN3LO} and~\ref{tab: scale_uncertainty_aN3LO-I} we present representative values for the NCDY and CCDY invariant-mass distribution and their scale dependence for various PDF sets for some representative bins.

\begin{table}
\begin{center}
\renewcommand{\arraystretch}{1.5}
\begin{tabular}{|c|c|c|c|c|}
\hline
Process  \rule{0pt}{3ex} & PDF & $\Sigma_{\textrm{N$^3$LO}}^{(5,0)}(\mu_F^0,\mu_R^0)$(pb) &\multicolumn{2}{c|}{$ \delta_\pm $}   \\
\hline
\multirow{4}{*}{NCDY }
 & {\tt {MSHT20nnlo\_as118}} & $\num{88.177009}$&$^{ + \num{1.022}}_{ - \num{1.533}}$ & $^{+ \num{1.16}\%}_{ - \num{1.74}\%}$ \\
 &&&&\\
 & {\tt {MSHT20an3lo\_as118}-I} & $\num{88.685649} $&$^{+ \num{1.018}}_{ - \num{1.609}}$ & $^{ + \num{1.15}\%}_{ - \num{1.81}\%}$ \\
 &&&&\\
 & {\tt {MSHT20an3lo\_as118}}-II & $\num{88.692056}$&$ ^{+ \num{1.018}} _{- \num{1.609}}$ & $^{ + \num{1.15}\%}_{ - \num{1.81}\%}$ \\
 \hline
\end{tabular}
\end{center}
\caption{\label{tab: scale_uncertainty_NNLO}
The central value and the scale dependence of the NCDY process at $\text{N}^3$LO in the 5FS for different PDF sets for the $Q$-bin (40 GeV, 50 GeV).
}
\end{table}

\begin{table}
\begin{center}
\renewcommand{\arraystretch}{1.5}
\begin{tabular}{|c|c|c|c|c|}
\hline
 Process & PDF & $\Sigma_{\textrm{N$^3$LO}}^{(5,+1)}(\mu_F^0,\mu_R^0)$(pb) &\multicolumn{2}{c|}{$ \delta_\pm $}   \\
\hline
\multirow{4}{*}{CCDY ($W^{+}$)}
 & {\tt {MSHT20nnlo\_as118}} & $\num{11170.036071}$&$^{ + \num{83.668}}_{ - \num{111.192}}$ & $^{+ \num{0.749}\%}_{ - \num{0.995}\%}$ \\
 &&&&\\
 & {\tt {MSHT20an3lo\_as118}-I}  & $\num{11255.129242}$&$^{ + \num{80.264}}_{ - \num{113.163}}$ & $^{ + \num{ 0.71 }\%}_{ - \num{1.01}\%}$ \\
 &&&&\\
 & {\tt {MSHT20an3lo\_as118}}-II  & $\num{11255.521004}$&$^{ + \num{80.266}}_{ - \num{113.165}}$ & $^{ + \num{0.713}\%}_{ - \num{1.005}\%}$ \\
 \hline
\end{tabular}
\end{center}
\caption{\label{tab: scale_uncertainty_aN3LO}
The central value and the scale dependence of the CCDY($W^+$) process at $\text{N}^3$LO in the 5FS for different PDF sets for the $Q$-bin (50 GeV, 150 GeV).
}
\end{table}

\begin{table}
\begin{center}
\renewcommand{\arraystretch}{1.5}
\begin{tabular}{|c|c|c|c|c|}
\hline
Process & PDF & $\Sigma_{\textrm{N$^3$LO}}^{(5,-1)}(\mu_F^0,\mu_R^0)$(pb) &\multicolumn{2}{c|}{$ \delta_\pm $}  \\
\hline
\multirow{4}{*}{CCDY ($W^{-}$)}
 &&&&\\
 & {\tt {MSHT20nnlo\_as118}} & $\num{8236.581705}$&$^{ + \num{64.391}}_{ - \num{86.055}}$ & $^{+ \num{0.782}\%}_{ -\num{1.045}\%}$ \\
 &&&&\\
 &  {\tt {MSHT20an3lo\_as118}-I}  & $\num{8310.569143} $&$^{+ \num{62.865}}_{ - \num{86.624}}$ & $^{+ \num{0.756}\%}_{ -\num{1.042}\%}$ \\
  &&&&\\
 & {\tt {MSHT20an3lo\_as118}}-II & $\num{8310.737783}$&$^{ + \num{62.866}}_{ - \num{86.624}}$ & $^{+ \num{0.756}\%}_{ -\num{1.042}\%}$ \\
 \hline
\end{tabular}
\end{center}
\caption{\label{tab: scale_uncertainty_aN3LO-I}
The central value and the scale dependence of the CCDY($W^-$) process at $\text{N}^3$LO in the 5FS for different PDF sets for the $Q$-bin (50 GeV, 150 GeV).
}
\end{table}

\subsection{PDF uncertainty}
PDFs are non-perturbative in nature. They cannot be computed from first principles using perturbation theory, but instead they need to be extracted from experimental measurements. This induces an uncertainty coming from the data as well as the methodology used in the fit. Each PDF set provides its own prescriptions to estimate the PDF uncertainty.
 The MSHT2022 PDF sets provide
$2n$ eigenvector PDF sets, $S_k^{\pm}~(k = 1, . . . , n)$. 
We calculate \emph{Hessian asymmetric PDF uncertainties}, denoted $\delta_{\pm}(\textrm{PDF})$, on the binned invariant-mass distribution as follows:
\beq\bsp \label{eq:pdfuncformula}
\delta_{+}(\textrm{PDF})^2 &=\sum\limits_{k=1}^{n}  \max \Big[ \Sigma^{(5,\kappa)}_{\textrm{N$^3$LO}}(S_k^{+}) - \Sigma^{(5,\kappa)}_{\textrm{N$^3$LO}}(S_k^{0}), \Sigma^{(5,\kappa)}_{\textrm{N$^3$LO}}(S_k^{-}) - \Sigma^{(5,\kappa)}_{\textrm{N$^3$LO}}(S_k^{0}), 0 \Big]^2\,, \\
\delta_{-}(\textrm{PDF})^2 &= \sum\limits_{k=1}^{n}  \max \big[ \Sigma^{(5,\kappa)}_{\textrm{N$^3$LO}}(S_k^{0})- \Sigma^{(5,\kappa)}_{\textrm{N$^3$LO}}(S_k^{+}), \Sigma^{(5,\kappa)}_{\textrm{N$^3$LO}}(S_k^{0}) - \Sigma^{(5,\kappa)}_{\textrm{N$^3$LO}}(S_k^{-}), 0 \big] ^2\,.
\esp\eeq
The uncertainty on the unbinned distribution is again computed analoguously.
Here, $n=52$ is used for the { \tt{MSHT20aN3LO\_as118}} Set I and $n=52$ with the{ \tt{MSHT20aN3LO\_as118}}Set II. We use PDF sets that incorporate approximate N$^3$LO QCD corrections, as described in~\cite{McGowan:2022nag}. For these theoretical PDF uncertainties, $n=32$ is used for { \tt{MSHT20aN3LO\_as118}} set I (and NNLO sets).  For the central values, we utilize the 0$^{\textrm{th}}$ member for each MSHT20 PDF sets. 

\begin{figure}
    \centering
    \includegraphics[width=0.7\linewidth]{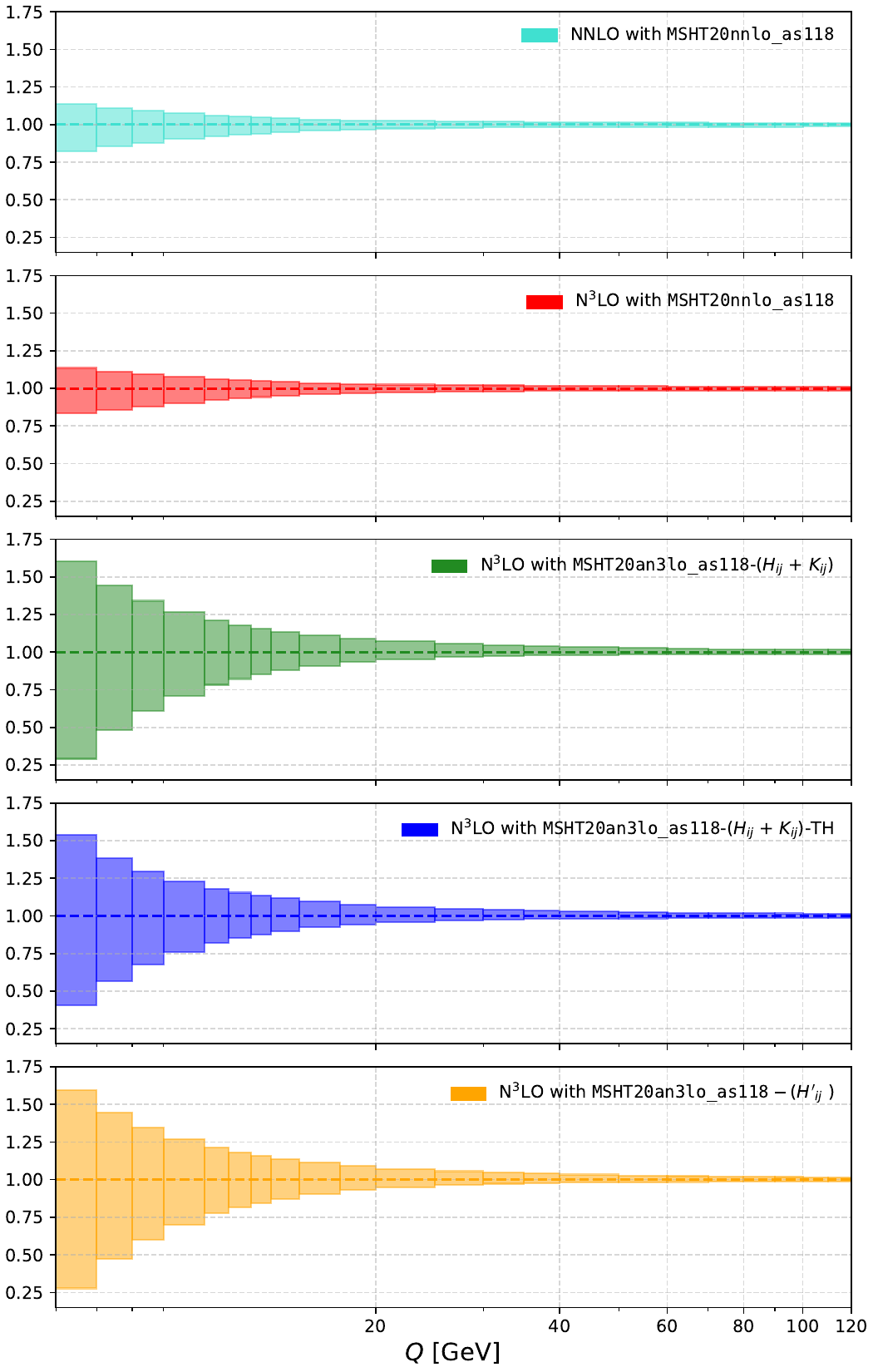}
    \caption{The PDF uncertainty for for the invariant-mass distribution for the NCDY process for various $Q$-bins. The bands are obtained using the maximum and minimum value of $\Sigma_{\textrm{N$^3$LO}}^{(5,0)}$ across the members for each PDF set. The $y$-axis corresponds to the ratio  of $\delta_\pm(\text{PDF})$ \emph{w.r.t.} the $0^{\text{th}}$ PDF member of each set.}
    \label{fig:newpdfun}
\end{figure}

Using the formula given in eq.~\eqref{eq:pdfuncformula}, we plot in fig.~\ref{fig:newpdfun} the PDF uncertainties as ratios with respect to the central value of the 0$^{\textrm{th}}$ PDF member in each PDF sets, as a function of the invariant mass $Q\in (4\textrm{ GeV}, 200\textrm{ GeV})$. The different PDF sets considered in this comparison are the NNLO and aN$^3$LO sets of MSHT20. In tables~\ref{tab: ggH_results_mh21}, \ref{tab: ggH_results_mh22} and~\ref{tab: ggH_results_mh23} we show representative PDF uncertainties for the bin $(40\textrm{ GeV}, 50\textrm{ GeV})$ for NCDY and $(50\textrm{ GeV}, 150\textrm{ GeV})$ for CCDY.
From these results, we observe an increase in the PDF uncertainties for the a$\text{N}^3$LO sets compared to the NNLO sets. This feature has also been noted in various other processes, cf.~\cite{McGowan:2022nag}. From the tables, we can also see that the PDF uncertainty is approximately 3\% for the aN$^3$LO sets, which is quite significant.

\begin{figure}
    \centering
    \includegraphics[width=1.0\linewidth]{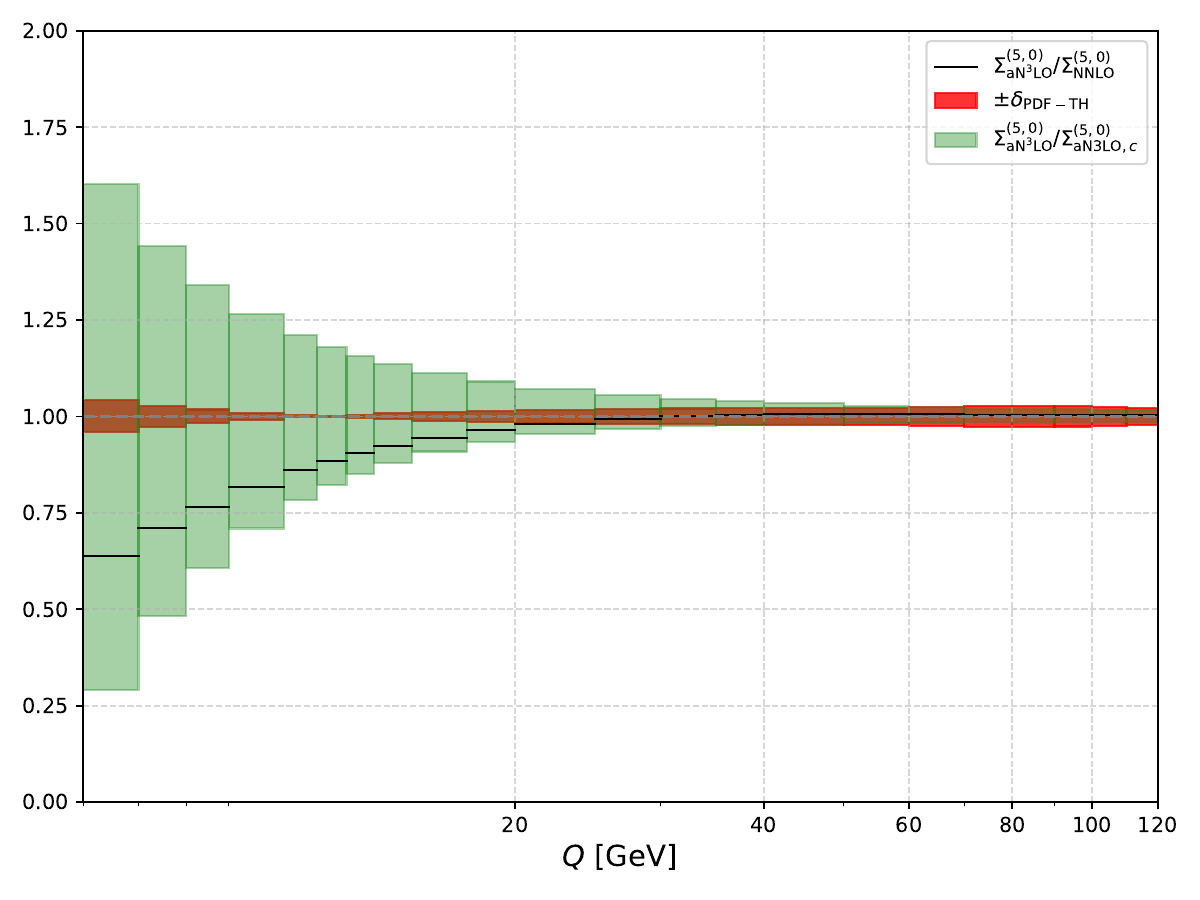}
    \caption{Comparison for the NCDY process of the PDF-TH uncertainty in eq.~\eqref{eq:newpdfthformula} with the {\tt{MSHT20nnlo\_as118}} and {\tt{MSHT20nlo\_as118}}  PDF sets (red band), and the central value (black dots) and the PDF uncertainty (green band) with the {\tt{MSHT20an3lo\_as118}} set. }
    \label{fig:newpdfth1}
\end{figure}

To further assess the uncertainties associated with the NNLO and aN$^3$LO PDF sets, we also present the results in fig.~\ref{fig:newpdfth1}. The $\text{N}^3$LO cross-section is plotted using two different PDF sets, namely  {\tt{MSHT20an3lo\_as118}} and {\tt{MSHT20nnlo\_as118}}. The black line in this plot shows the increment of the $\text{N}^3$LO cross-section  using {\tt{MSHT20an3lo\_as118}} with respect to the {\tt{MSHT20nnlo\_as118}} PDF. The red bands represent the theoretical uncertainty arising from the PDFs for the NNLO cross-section, as in~\cite{Anastasiou:2016cez}, calculated according to:
\begin{align}\label{eq:newpdfthformula}
    \delta(\text{PDF-TH}) = \frac{1}{2} \frac{|\Sigma^{(5,0)}_{\text{NNLO}}( \text{NNLO PDF})- \Sigma^{(5,0)}_{\text{NNLO}} (\text{NLO PDF})|}{\Sigma^{(5,0)}_{\text{NNLO}} (\text{NNLO PDF})}.
\end{align}
We observe that for $Q$ above 40 GeV, the uncertainty band obtained from $\delta(\text{PDF-TH})$ contains the aN$^3$LO central value and PDF uncertainty bands. For small values of $Q$, we see that the central value departs from the $\delta(\text{PDF-TH})$ band. At the same time, the PDF uncertainty computed at aN$^3$LO increases dramatically, indicating that further knowledge about N$^3$LO PDFs is required.

\begin{table}

\begin{center}
\renewcommand{\arraystretch}{1.5}
\begin{tabular}{|c|c|c|c|c|}
\hline
Process & PDF order  & $\Sigma_{\textrm{N$^3$LO}}^{(5,0)}$(pb)  & \multicolumn{2}{c|}{$\delta_{\pm}$(PDF)} \\
\hline
\multirow{4}{*}{NCDY}
 & aN$^{3}$LO (no theory unc.) & $\num{88.685649}$ & $^{+\num{2.589380}} _{- \num{1.656312}}$ & $ ^{+\num{2.92}\% }_{- \num{1.87}\%}$  \\
 &&&&\\
 & aN$^{3}$LO ($H_{ij} + K_{ij}$) & $\num{88.685649}$ & $^{+\num{3.006931}}_{- \num{1.726206}}$ & $^{+ \num{3.39}\%}_{ - \num{1.95}\%}$ \\
 &&&&\\
	& aN$^{3}$LO ($H_{ij}^{\prime}$) & $\num{88.692056}$ & $^{+\num{3.016278}}_{ - \num{1.768223}}$ & $^{ + \num{3.40}\%}_{ -\num{1.99}\%}$ \\
    &&&&\\
 & NNLO & $\num{88.25309135185499}$ & $^{+ \num{1.49669664513119}} _{- \num{1.4870120112941894}}$  & $^{ + \num{1.70}\%} _{- \num{1.68}\%}$ \\
 \hline
\end{tabular}
\caption{\label{tab: ggH_results_mh21} The PDF uncertainty for the NCDY process at for $Q$-bin (40 GeV, 50 GeV).}
\end{center}
\end{table}

\begin{table}
\begin{center}
\renewcommand{\arraystretch}{1.5}
\begin{tabular}{|c|c|c|c|c|}
\hline
Process& PDF order & $\Sigma_{\textrm{N$^3$LO}}^{(5,+1)}$(pb) &  \multicolumn{2}{c|}{$\delta_{\pm}$(PDF)}  \\
\hline
\multirow{4}{*}{CCDY($W^+$)}
 & aN$^{3}$LO (no theory unc.) & $\num{8310.56914334783}$&$^{ + \num{146.21300585289637}}_{ - \num{124.35986834783779}}$ & $^{ + \num{1.7593621246738818}\%}_{ - \num{1.4964061570606302}\%}$ \\
 &&&&\\
 & aN$^{3}$LO ($H_{ij} + K_{ij}$) & $\num{8310.56914334783}$&$ ^{+ \num{ 149.38544220830158}}_{ -\num{126.56585881474487}}$ & $^{ + \num{1.7975356396363866}\%}_{ - \num{1.5229505540670958}\%}$ \\
 &&&&\\
 & aN$^{3}$LO ($H_{ij}^{\prime}$) & $\num{8310.737783162844}$&$^{ + \num{ 146.57431360492416}}_{ - \num{125.64136105428607}}$ & $^{+ \num{1.7636739051240036}\%}_{ -\num{1.5117955148197484}\%}$ \\
 &&&&\\
 & NNLO & $\num{8236.581705451914}$&$ ^{+ \num{113.3871187761268}}_{ - \num{135.5851428666368}}$  & $^{ + \num{1.3766283493682119}\%}_{ - \num{1.646133647613682}\%}$ \\
 \hline
\end{tabular}
\caption{\label{tab: ggH_results_mh22}The PDF uncertainty for the CCDY($W^+$) process at for $Q$-bin (50 GeV, 150 GeV).}
\end{center}
\end{table}

\begin{table}
\begin{center}
\renewcommand{\arraystretch}{1.5}
\begin{tabular}{|c|c|c|c|c|}
\hline
Process & PDF order & $\Sigma_{\textrm{N$^3$LO}}^{(5,-1)}$(pb) & \multicolumn{2}{c|}{$\delta_{\pm}$(PDF)} \\
\hline
\multirow{4}{*}{CCDY($W^-$)}
 & aN$^{3}$LO (no theory unc.) & $\num{11255.12924165649}$&$ ^{+ \num{217.2628971671925}}_{ - \num{171.53818197180468}}$ & $^{\num{1.9303456451044407}\%}_{ - \num{1.524088957920827}\%}$ \\
 &&&&\\
 & aN$^{3}$LO ($H_{ij} + K_{ij}$) & $\num{11255.12924165649}$&$^{ + \num{222.63795645000414}}_{ - \num{173.12780175203054}}$ & $^{+ \num{1.9781021760816055}\%}_{ - \num{1.5382124721523873}\%}$ \\
 &&&&\\
 & aN$^{3}$LO ($H_{ij}^{\prime}$) & $\num{11255.521004186627}$&$ ^{+ \num{217.85573612195475}}_{ - \num{169.67544193072052}}$ & $^{+ \num{ 1.9355455517423021}\%}_{ -\num{1.5074863426367175}\%}$ \\
 &&&&\\
 & NNLO & $\num{11170.03607089626}$&$ ^{+ \num{163.9756234932771}}_{ - \num{167.8154414008224}}$  & $^{+ \num{1.467995469777566}\%}_{ - \num{1.502371526248413}\%}$ \\
 \hline
\end{tabular}
\end{center}
\caption{\label{tab: ggH_results_mh23}The PDF uncertainty for the CCDY($W^-$) process at for $Q$-bin (50 GeV, 150 GeV).}
\end{table}

The PDF uncertainty is closely connected to the uncertainty due to the strong coupling constant $\alpha_s$, which also needs to be extracted from experiment, and is also often determined from a global fit together with the PDFs. In order to determine the $\alpha_s$-uncertainty, we use the values of the strong coupling provided by {\tt{MSHTNNLO\_as\_smallrange}}. We then fit our predictions for the invariant-mass distribution using a {\tt{GNU}} fitting procedure to fit a function of the type of $f(x) = a + b ~ x + c ~ x^2 + d ~x^3$. The parameters $\{a,b,c,d\}$ are determined using a non-linear least square fit using the values of the invariant-mass distribution at $\alpha_s(M_z^2) = 0.118$ given by {\tt{MSHTNNLO\_as\_smallrange}}. We note that these data are only available for the NNLO PDF sets, and so we limit the study of the $\alpha_s$-uncertainty to this case. Then with the extracted values of the parameters we make an estimate of the interpolated bin values at $\alpha_s=0.1171$ and $0.1189$, which corresponds to the PDG world average~\cite{ParticleDataGroup:2024cfk}.
The uncertainty estimation  with respect to the deviation from the central value of 0.118 is calculated as:
\beq\bsp
\delta_{\pm}(\alpha_s) &= {\Sigma_{\textrm{NNLO}}^{(5,\kappa)}(\alpha_s=0.118 \pm 0.0009) - \Sigma_{\textrm{NNLO}}^{(5,\kappa)}(\alpha_s=0.118)}\,  .
\esp\eeq
In the following we discuss the $\alpha_s$-uncertainty independently from the PDF uncertainty. 
We note, however, that, following the recommendation by~{{\tt PDF4LHC}}~\cite{Butterworth:2015oua}, the PDF and $\alpha_s$ uncertainties should be added in quadrature,
\beq
\delta_{\pm}(\textrm{PDF}+\alpha_s) = \sqrt{\delta_{\pm}(\textrm{PDF})^2 +\delta_{\pm}(\alpha_s)^2 }\,.
\eeq

\begin{figure}
    \centering
\includegraphics[width=1.0\linewidth]{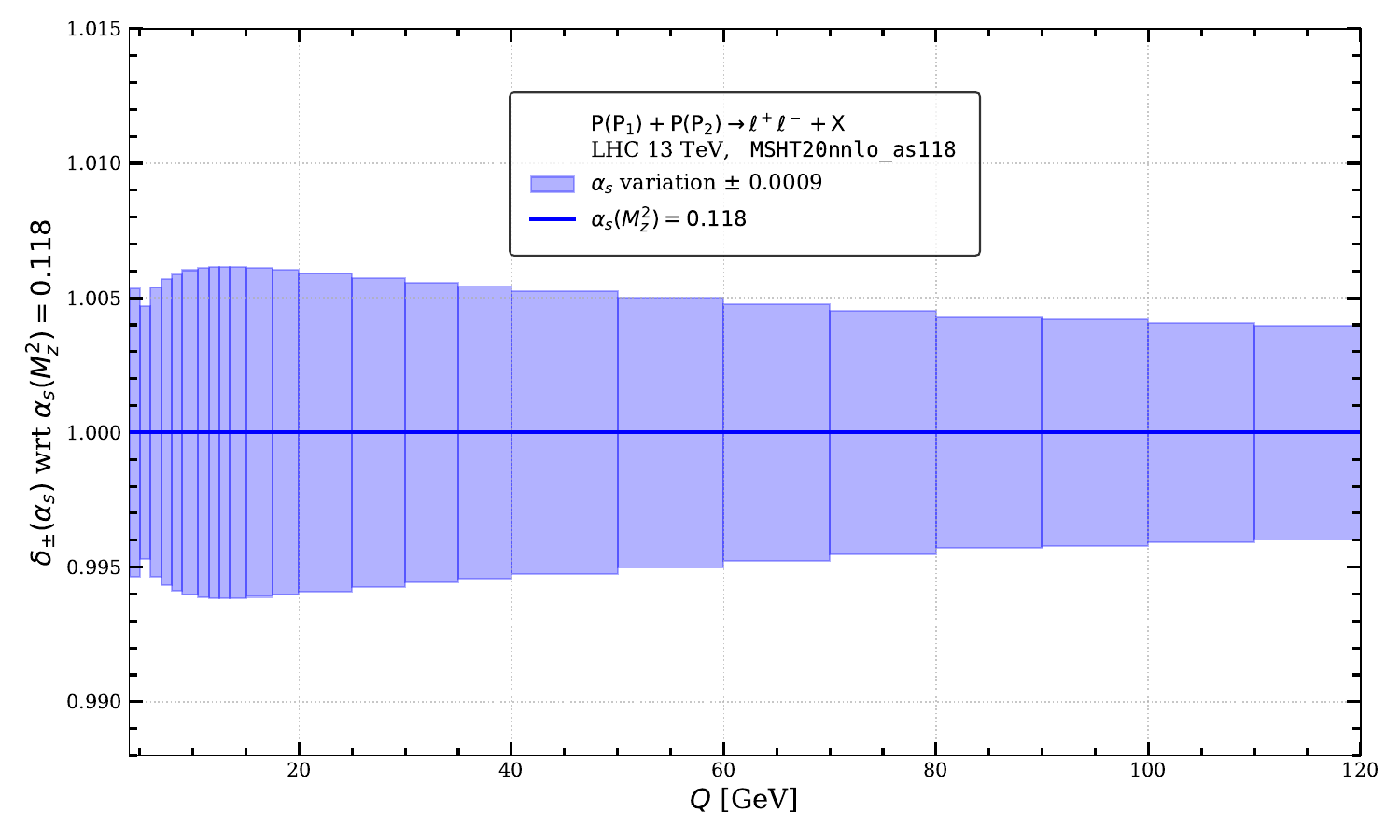}
    \caption{Ratio plot for the $\alpha_s$ variation \emph{w.r.t.} $\Sigma_{\textrm{NNLO}}^{(5,0)}(\alpha_s=0.118)$.}
    \label{fig:newasun}
\end{figure}

In fig.~\ref{fig:newasun} we show the $\alpha_s$-uncertainty on the invariant-mass distribution for the NCDY process across bins ranging from 4 to 120 GeV. We show the ratio of the uncertainty in $\alpha_s$ relative to its central value $\alpha_s=0.118$. In table~\ref{tab: alphas_uncertainty} we also present the uncertainties for the NCDY and CCDY processes for selected bins.
We find that the uncertainties due to the above variations are much less as compared to the PDF uncertainties. 

\begin{table}
\begin{center}
\renewcommand{\arraystretch}{1.8}
\begin{tabular}{|c|c|c|c|}
\hline
 Process & $\Sigma_{\textrm{NNLO}}^{(5,\kappa)}$(pb)&  \multicolumn{2}{c|}{$\delta_{\pm}(\alpha_s)$} \\
 \hline
   NCDY (40-50 GeV) & $\num{91.5753528346672} $&$^{+ \num{0.0104807028509481}}_{ - \num{0.0103741419276474}}$ & $^{+ \num{1.04807028509}\%}_{ - \num{1.037}\%}$ \\
 &&&\\
  CCDY ($W^{+}$) (50-150 GeV) & $\num{11402.832056351}$&$^{ + \num{0.00833523717229781}}_{ - \num{0.00391692353556233}}$ & $^{+ \num{0.833523717229781}\%}_{ - \num{0.391692353556233}\%}$ \\
  &&&\\
  CCDY ($W^{-}$) (50-150 GeV) & $\num{8457.89791593468}$&$ ^{+ \num{0.00819544743482936}}_{ - \num{0.00814116022257895}}$ & $^{+ \num{0.819544743482936}\%}_{ -\num{0.814116022257895}\%}$ \\
 \hline
\end{tabular}
\end{center}
\caption{\label{tab: alphas_uncertainty}The $\alpha_s$ uncertainty at NNLO for all the processes for selected $Q$-bins.}
\end{table}

\subsection{Mass effects in DY processes}
In this section we discuss the impact of the matching of the 4FS and 5FS using the MVFNS described in section~\ref{sec:matching_theory}. We limit this study to NNLO, i.e. up to $\mathcal{O}(\alpha_s^2)$, due to the absence of the complete set of relevant two- and three-loop computations needed for the N$^3$LO contributions in the 4FS. Throughout this computation we use {\tt{MSHT20nnlo\_as118}} PDF sets with the default quark mass values given in eq.~\eqref{eq:quarkmassvalues}. We consider predictions for invariant-mass bins matched using the MVFNS according to
\beq
\Sigma_{\textrm{NNLO}}^{(\textrm{mat.},\kappa)} = \Sigma_{\textrm{NNLO}}^{(5,\kappa)} + \sum_{f=\{c,b\}}\sum_{l=0}^{2} a_s(\mu_R)^l\Sigma_{pc,[f]}^{(5,\kappa,l)}\,,
\eeq
where the power-corrections $\Sigma_{pc,[f]}^{(5,\kappa,l)}$ were defined in eq.~\eqref{eq:Sigma_power_corrections}. For NCDY it starts with $l=2$ and for CCDY it starts with $l=1$.

\begin{table}
\begin{center}
\renewcommand{\arraystretch}{1.8}
\begin{tabular}{|c|c|c|c|}
\hline
Process & $\Sigma^{(n_f,0)}(Q_{min},Q_{max})$              & Prediction (pb)   \\ \hline
 &$\Sigma_{\text{N}^2\text{LO}}^{(5,0)}(80,105)$            & $1824.63$\\ \cline{2-3}
NCDY &$\Sigma_{pc,[b]}^{(5,0,2)}(80,105)$          & $0.965396$       \\ \cline{2-3}
&$\Sigma_{pc,[c]}^{(5,0,2)}(80,105)$                        & $0.088011$                \\ \hline
&$\Sigma_{\text{N}\text{LO}}^{(5,+1)}(50,150)$            & $11482.76$\\ \cline{2-3}
CCDY &$\Sigma_{pc,[c]}^{(5,+1,1)}(50,150)$                        & $0.662841$                \\ \cline{2-3}
 &$\Sigma_{pc,[c]}^{(5,+1,2)}(50,150)$                        & $0.0101773$                \\ \hline
 &$\Sigma_{\text{N}\text{LO}}^{(5,-1)}(50,150)$            & $8524.48$\\ \cline{2-3}
CCDY &$\Sigma_{pc,[c]}^{(5,-1,1)}(50,150)$                        & $0.662637$                \\ \cline{2-3}
 &$\Sigma_{pc,[c]}^{(5,-1,2)}(50,150)$                        & $0.0104385$\\
 \hline
\end{tabular}
\end{center}
	\caption{Predictions from the 5FS and the massive power-corrections at  
 are shown for the central scale $\mu_R=\mu_F=Q$ for physical values of quark masses in eq.~\eqref{eq:quarkmassvalues}.}
\label{Table:Fiducial}
\end{table}

\begin{figure}
    \centering
    \includegraphics[width=1.0\linewidth]{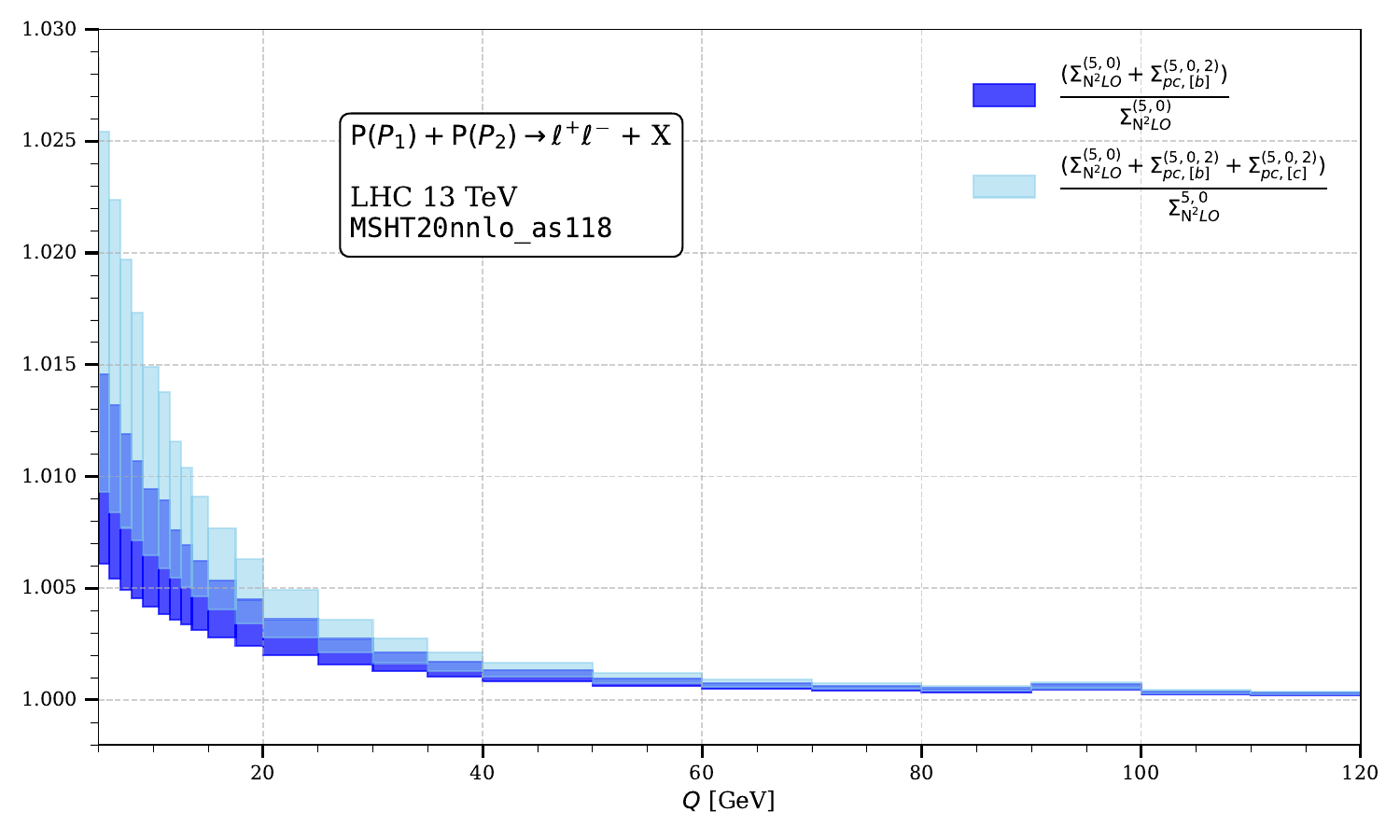}
    \caption{Impact of the power corrections on the invariant-mass distribution of the NCDY process. Predictions are obtained with the PDF set {\texttt{MSHTnnlo\_as118}}.}
    \label{fig:sigma_extreme_PC1}
\end{figure}

In fig.~\ref{fig:sigma_extreme_PC1} we show the ratio of the NCDY invariant-mass distribution at NNLO computed in a MVFNS and the 5FS. We show the results for the cases where only the bottom quark or both the bottom and charm quarks are treated as massive. We observe that, as expected, for invariant masses above 40 GeV, the effect of the power-suppressed terms in the quark masses are at the permille level, and so they can safely be neglected compared to the other sources of uncertainties we have considered. In table~\ref{Table:Fiducial} we show the effect of the power-corrections on the invariant-mass bin (80 GeV, 105 GeV) around the $Z$ boson mass, and we see that the effect is negligible. For small invariant masses the effect of the power corrections increases, reaching the percent level around 10 GeV. We also observe that the inclusion of the charm quark masses leads to a larger effect than the inclusion of the bottom quark mass alone. 

For CCDY, the effects of power corrections at NLO are shown in table \ref{Table:Fiducial} for $W^\pm$ production for NNLO it is shown in tables \ref{Table:Fiducial-CDYas21} and \ref{Table:Fiducial-CDYas22}. The massive corrections were obtained using {\tt{MCFM-10.3}}~\cite{Campbell:2015qma,Campbell:2019dru} and ${\rm{d}}\sigma_{\ln[m]}^{(5,0,k)}$ and ${\rm{d}}\sigma_{n_f}^{(5,0,k)}$ were computed using in-house codes. 
We faced numerical challenges for the massive computations in the low-mass region. 
The corresponding error could be brought down to sub-permille level. We see that the power corrections for CCDY are very small as compared to NCDY. 

We conclude that for precision studies at the percent level in the small invariant-mass region, the effects of bottom and charm masses should not be neglected. We note that the LHCb experiment performs measurements of DY processes for invariant masses down to 4 GeV.

\begin{figure}
    \centering
          \includegraphics[width=1.0\linewidth]{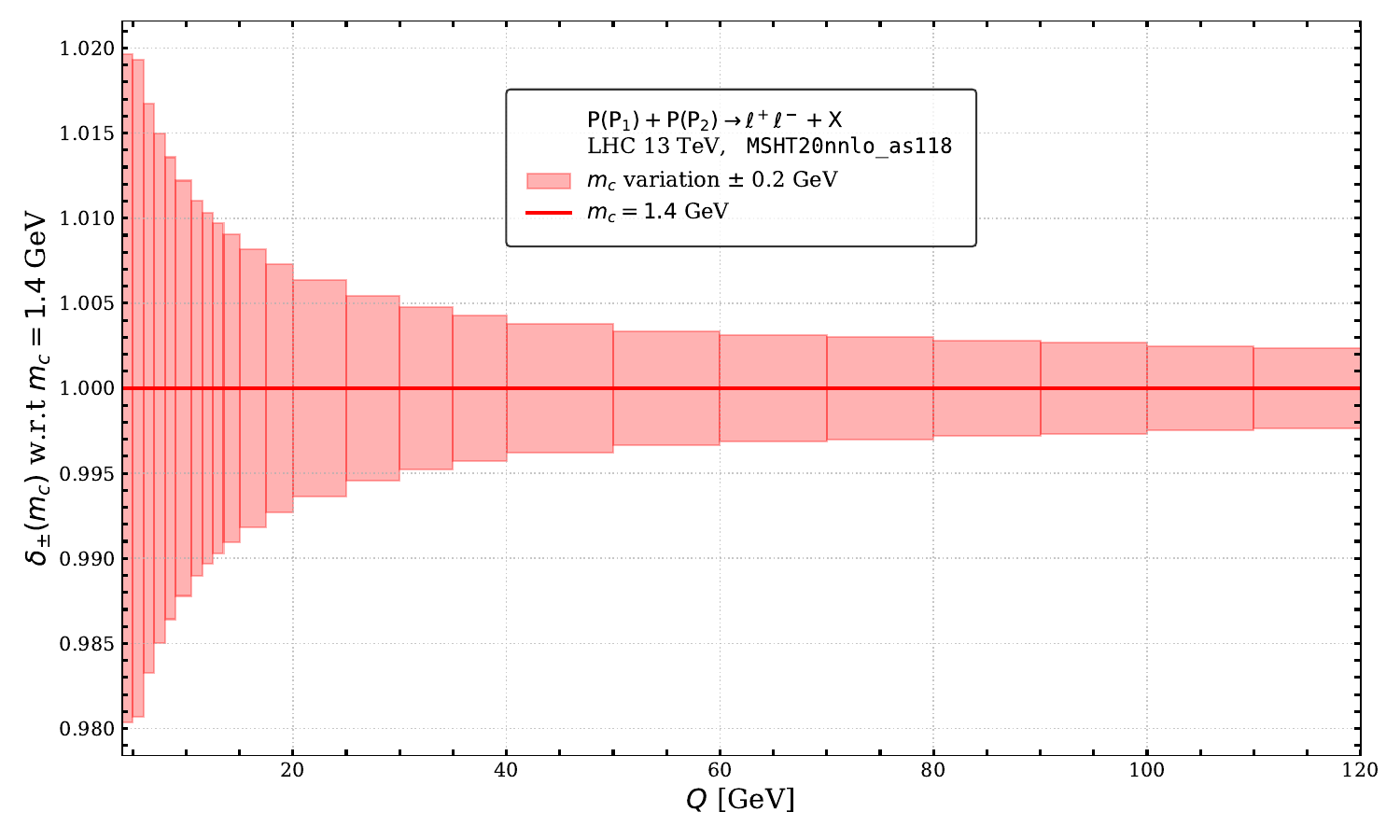}
\caption{Impact of varying the charm quark masses on the  NNLO invariant-mass distribution for the NCDY process.}
    \label{fig:new3}
\end{figure}

We have also assessed the uncertainty on the invariant-mass distribution induced by varying the values of the quark masses. To do so, we use a similar approach for the determination of the $\alpha_s$-uncertainty, i.e., we use the PDF sets {\tt{MSHTNNLO\_mcrange\_nf5}} and {\tt{MSHTNNLO\_mbrange\_nf5}} for varying charm and bottom quark masses. The resulting uncertainties on the invariant-masses bins are then computed as
\beq\bsp
\delta_{\pm}(m_c) &\,= \Sigma_{\textrm{NNLO}}^{(\textrm{mat.},\kappa)}(m_c=1.4\pm0.2 \textrm{ GeV})-\Sigma_{\textrm{NNLO}}^{(\textrm{mat.},\kappa)}(m_c=1.4 \textrm{ GeV})\,,\\
\delta_{\pm}(m_b) &\,= \Sigma_{\textrm{NNLO}}^{(\textrm{mat.},\kappa)}(m_b=4.75\pm0.5 \textrm{ GeV})-\Sigma_{\textrm{NNLO}}^{(\textrm{mat.},\kappa)}(m_c=4.75 \textrm{ GeV})\,.
\esp\eeq
In fig.~\ref{fig:new3} and tables~\ref{tab: m_b_variation} and~\ref{tab: m_c_variation} we show the impact of varying the quark masses. We observe that in all cases the impact is below the percent level, with the charm quark having overall a larger effect than the bottom quark. In fact, the uncertainties arising from the variation of the bottom mass in NCDY is below the sub permille level, as can be seen in table~\ref{tab: m_b_variation}. 
Instead, the uncertainty coming from the variation of $m_c$ is as significant as for the uncertainty from the $\alpha_s$ variation in the low-Q region for NCDY, which can be seen in fig.~\ref{fig:new3} and table~\ref{tab: m_c_variation}.

\begin{table}
\begin{center}
\renewcommand{\arraystretch}{1.5}
\begin{tabular}{|c|c|c|c|c|}
\hline
 Process & $Q$-bin  & $\Sigma_{\textrm{NNLO}}^{(\textrm{mat.},\kappa)} $ (pb)&\multicolumn{2}{c|}{$ \delta_{\pm}(m_b)$} \\
\hline
 NCDY& (40 GeV, 50 GeV) & $\num{91.5909182417414}$&$^{ + \num{0.000544125105232647} }_{- \num{0.000646170890438374}}$ & $^{+ \num{0.05}\%}_{ - \num{0.06}\%}$ \\
 &&&&\\
  CCDY ($W^{+}$)& (50 GeV, 150 GeV) & $\num{11452.9674962526} $&$^{+ \num{0.00190769320071866}}_{ - \num{0.00229670919299895}}$ & $^{+ \num{0.190769320071866}\%}_{ - \num{0.229670919299895}\%}$ \\
 &&&&\\
  CCDY ($W^{-}$)& (50 GeV, 150 GeV) & $\num{8460.35286363186}$&$^{ + \num{0.00180018950391709}}_{ - \num{0.00227792515448087}} $ & $^{+ \num{0.180018950391709}\%}_{ -\num{0.227792515448087}\%}$ \\
 \hline
\end{tabular}
\end{center}
\caption{\label{tab: m_b_variation}Variation of the invariant-mass distributions at NNLO with the bottom $m_b$. The $m_b$ variation corresponds to $\pm 0.5$ GeV, and the central $m_b$ value  is 4.75 GeV. }
\end{table}

\begin{table}
\begin{center}
\renewcommand{\arraystretch}{1.5}
\begin{tabular}{|c|c|c|c|c|}
\hline
Process & $Q$-bin & $\Sigma_{\textrm{NNLO}}^{(\textrm{mat.},\kappa)} $ (pb)&\multicolumn{2}{c|}{$ \delta_{\pm}(m_c)$} \\
\hline
  NCDY& (40 GeV, 50 GeV) & $\num{91.5965382172613}$&$^{ + \num{0.00581898132700543}}_{ - \num{0.00762324222578774}}$ & $^{ + \num{0.58}\%}_{ - \num{0.76}\%}$ \\
 &&&&\\
  CCDY ($W^{+}$) &(50 GeV, 150 GeV) & $\num{11451.4057151254}$&$^{ + \num{0.00310489906389174}}_{ - \num{0.00321085888987096}}$ & $^{ + \num{0.310489906389174}\%}_{ - \num{0.321085888987096}\%}$ \\
 &&&&\\
  CCDY ($W^{-}$) &(50 GeV, 150 GeV) & $\num{8459.15196052564}$&$^{ + \num{0.0017319343979214}}_{ - \num{0.00283790923390715} }$ & $^{+ \num{0.17319343979214}\%}_{ -\num{0.283790923390715}\%}$ \\
 \hline
\end{tabular}
\end{center}
\caption{\label{tab: m_c_variation}
Variation of the invariant-mass distributions at NNLO with the bottom $m_c$. The $m_c$ variation corresponds to $\pm 0.2$ GeV, and the central $m_c$ value  is 1.4 GeV.}
\end{table}

\section{Conclusions}
\label{sec:conclusions}
\begin{figure}
    \centering
    \includegraphics[width=1.0\linewidth]{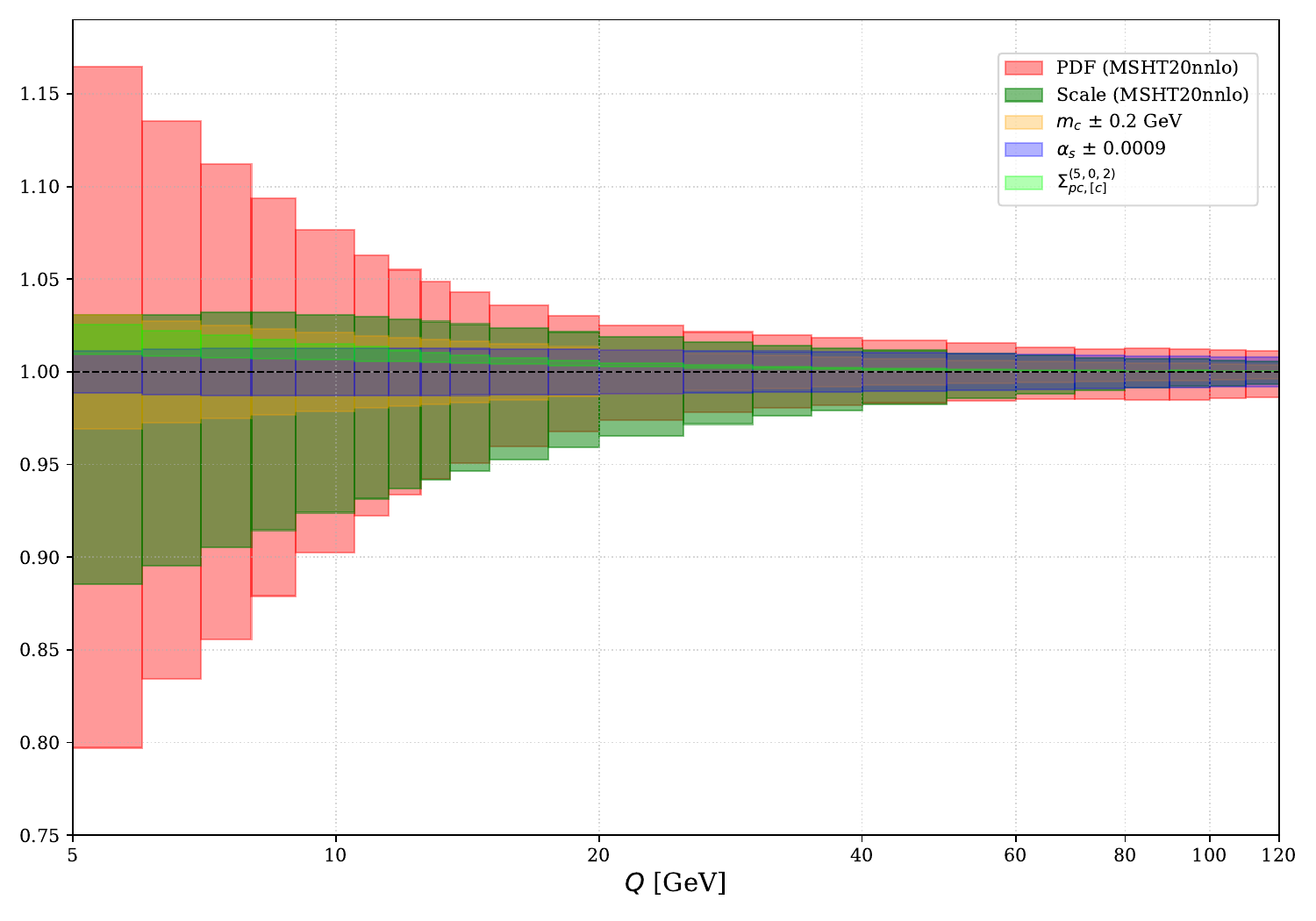}
    \caption{The hierarchy of all sources of uncertainty present in NCDY. Predictions are obtained with the PDF set {\texttt{MSHTnnlo\_as118}}. The $y$-axis presents the variation of the cross-section \emph{w.r.t.} to scale, PDF, power correction, $\alpha_s$, and mass variations for all uncertainties. }
    \label{fig:All_unc1}
\end{figure}
In this article we have performed a detailed analysis of both NCDY and CCDY processes in QCD perturbation theory, including various sources of uncertainty.
In each case the study has been performed within the vicinity of the $Z$ and $W^{\pm}$ gauge boson masses respectively, while for the NCDY process the analysis has also been extended to the region of low $Q$ (i.e. low invariant dilepton masses).

To extend the calculations to the low-$Q$ region, we have included the exact effect of charm and bottom quark masses up to $\mathcal{O}(\alpha_s^2)$. 
The method has also been applied to the case of CCDY, which is to our knowledge, the first example of the matching for the CC process.
The details and numerical validation of our implementation to calculate these effects has also been provided.

The evaluation of PDF uncertainties obtained from the MSHT20 PDF sets using N$^3$LO and NNLO grids reveals that the PDF uncertainties increase for the N$^3$LO sets as compared to the NNLO sets, an effect that is significant for low-$Q$ values. This feature has also been observed in other processes~\cite{McGowan:2022nag}.
Overall it is observed that the scale uncertainties arising from the N$^3$LO calculation are sub-dominant as compared to those from the PDFs.
In particular for $Q\sim M_W (M_Z)$ the PDF uncertainties are typically 2\%(3\%), and larger than the scale uncertainties.
The impact of uncertainties due to the choice of the charm quark mass in the global fit, as well as the uncertainty due to knowledge of the value of $\alpha_s(m_Z)$, introduces an uncertainty at the several per-mille level (see also~\cite{Cridge:2021qfd}).
The impact of power corrections due to charm and bottom quarks has a negligible (per-mille or below) effect on the total cross-section at $Q\sim M_W (M_Z)$.

In the low-$Q$ region the hierarchy of each of these physics effects does not significantly change, but their overall relative impact on the total cross-section does increase.
A summary of these effects is shown in fig.~\ref{fig:All_unc1} which details the size of each of these different sources of uncertainty.
In particular, the impact of uncertainties due to the charm quark mass in the global fit, the impact of power corrections from mass effects, and uncertainties related to knowledge of $\alpha_s$ each impact the cross-section at the percent level for $Q\sim 15$ GeV.
The scale uncertainties in this region are approximately $+(-)3(5)\%$ which means that the aforementioned sources of uncertainty are not negligible.
%
The scale uncertainty obtained with the aN$^{3}$LO ($H_{ij}^{\prime}$) PDF set was observed to be relatively large as compared to other such PDF sets~\cite{NNPDF:2024nan}, a feature that was rigorously checked using approximate NNPDF N$^3$LO sets.

Overall, this study provides a detailed assessment of the uncertainties and corrections that affect the Drell-Yan process predictions, laying the groundwork for future precision measurements and comparisons with experimental data.
Such data is critical to improve our understanding of hadron structure at small $x$, and has implications for the study of high-energy astrophysical processes.

%

\section*{Acknowledgments}

We thank Tobias Neumann of the {\tt MCFM} collaboration for his assistance with running the code in the low-mass region.  The work of P.M.\ was supported by the ERC Advanced Grant 101095857 Conformal-EIC. The work of E.C.\ was funded by the European Union (ERC Consolidator Grant LoCoMotive 101043686). Views and opinions expressed are however those of the author(s) only and do not necessarily reflect those of the European Union or the European Research Council. Neither the European Union nor the granting authority can be held responsible for them.

\appendix


\appendix

\section{Matching 3 vs 5 flavours}
\label{appendix:3vs5}
In section~\ref{sec:matching_theory} we have been considering a single heavy flavour quark at a time. The ratio of masses of charm and bottom quarks are of $\mathcal{O}(0.1)$, so it is difficult to assume the mass of charm to be zero at scales of the order of the bottom quark mass. Since it is possible to decouple the masses of both the quarks, we now consider the effects of $m_c$ and $m_b$ on the differential cross-section, with $m_c$ being the charm quark mass. Therefore, the logarithmic corrections will contain both $m_b$ and $m_c$ dependent terms. Extending eq.~\eqref{eq:bpdf5}, we derive the relation between PDFs of 3 flavours to those at 5 flavours~\cite{Blumlein:2018jfm}: 
\begin{align}\label{eq:bto5_exp}
f_b^{5} = f_{\bar{b}}^{5}= &a_s\,A_{Qg}^{(1,b)}\otimes f_g^{5}+a_s^2\,\Bigg[\bigg( A_{Qg}^{(2,b)} + \frac{1}{2}A_{Qg}^{(2,cb)}-\frac{2}{3}\bigg[\ln{\frac{m_c^2}{\mu^2}}+\ln{\frac{m_b^2}{\mu^2}}\bigg]A_{Qg}^{(1,b)}\bigg)\otimes f_g^{5} \nonumber\\&
+ \sum_{\substack{i=-3\\ i\neq 0}}^3A_{Q i}^{(2,b)}\otimes f_i^{5}\Bigg]+\mathcal{O}(a_s^3)\, , \nonumber \\  
f_c^{5} = f_{\bar{c}}^{5}=& a_s\,A_{Qg}^{(1,c)}\otimes f_g^{5} +a_s^2\,\Bigg[\bigg( A_{Qg}^{(2,c)} + \frac{1}{2}A_{Qg}^{(2,cb)}-\frac{2}{3}\bigg[\ln{\frac{m_c^2}{\mu^2}}+\ln{\frac{m_b^2}{\mu^2}}\bigg]A_{Qg}^{(1,c)}\bigg)\otimes f_g^{5}\nonumber\\&
+ \sum_{\substack{i=-3\\ i\neq 0}}^3 A_{Q i}^{(2,c)})\otimes f_i^{5}\Bigg]+\mathcal{O}(a_s^3)\,, 
\end{align}
where the OMEs $A_{ij}^{(k,Q)}$ and $A_{ij}^{(k,Q_1 Q_2)}$ depend on $\ln m_Q$ and $\ln m_{Q_1} \ln m_{Q_2}$, respectively,
\beq\bsp
    A_{Qg}^{(2,cb)} 
    = &- \frac{8}{3}\bigg( z^2 + (1-z)^2 \bigg)\ln{\frac{m_c^2}{\mu^2}}\ln{\frac{m_b^2}{\mu^2}}, \\
    A_{Qg}^{(1,Q)} &= -2 \bigg( z^2 + (1-z)^2 \bigg)\ln{\frac{m_Q^2}{\mu^2}}.
    \label{eq:defAQg}
\esp\eeq

Note that in eq.~\eqref{eq:bto5_exp} the contributions coming from the mixing of both the quark masses ($A_{Qg}^{2 ,cb}$), eventually gets cancelled with the product of $\ln \frac{m_Q}{\mu}$ and single-mass OME $(A_{Qg}^{1,Q})$. Therefore it is equivalent to single-mass OME expressions in  eq.~\eqref{eq:bpdf5}. The whole matching procedure to $a_s^2$ then becomes equivalent to adding the power corrections.

Similar generalisations of eq.~\eqref{eq:powerexpansionofAs} to 5 vs 3 matching terms for NCDY are as follows:

 \begin{align}
\delta \eta^{(5,2)} _{gg} =& \sum_{i\in \{c,b \}} 2 A_{ig }^{(1)} \otimes A_{ig}^{(1)} \otimes \eta_{q \bar{q}}^{(5,0)} + 4 A_{ig}^{(1)} \otimes \eta _{q g}^{(5,1)}.
 \label{eq:powerexpansionofAs5vs3}
\end{align}
 In eq.~\eqref{eq:powerexpansionofAs5vs3}, we have only included the contributions from $gg$-channels because the OME are still unavailable in the other channels in $z$-space and $gq$ and $ q \bar{q}$ channels are not present. 

As was done for the NCDY process, for the CCDY process we also replaced  the charm and bottom PDFs in all the partonic subprocesses using eq.~\eqref{eq:bto5_exp} to obtain the matching coefficients $\delta\eta_{ij}^{\pm(n_f, k)} $ as follows:
\begin{align}
    &\delta\eta_{g\bar u}^{-(5, 1)} = A_{Qg}^{(1,b)}\otimes \eta_{b\bar u}^{-(5,0)},\nonumber\\&
    \delta\eta_{gi}^{-(5,1)} = A_{Qg}^{(1,c)}\otimes \eta_{\bar c i}^{-(5,0)} , \qquad i \in \{ d,s\}\nonumber\\&
     \delta\eta_{qd}^{-(5,2)} = A_{Qq}^{(2,c)}\otimes \eta_{d\bar c }^{-(5,0)} , \nonumber\\&
     \delta\eta_{qs}^{-(5,2)} = A_{Qq}^{(2,c)}\otimes \eta_{s\bar c }^{-(5,0)} , \nonumber\\&
     \delta\eta_{g d}^{(2)} =  A_{Qg}^{(2,c)}\otimes \eta_{\bar c d }^{-(5,0)} + A_{Qg}^{(1,c)}\otimes \eta_{\bar c d }^{-(5,1)},  \nonumber\\&
     \delta\eta_{gs}^{-(5,2)} =  A_{Qg}^{(2,c)}\otimes \eta_{\bar c s }^{-(5,0)} + A_{Qg}^{(1,c)}\otimes \eta_{\bar c s }^{-(5,1)},  \nonumber\\&
     \delta\eta_{gg}^{-(5,2)} =  2 A_{Qg}^{(1,b)}\otimes \eta_{gb }^{-(5,1)} +2 A_{Qg}^{(1,c)}\otimes \eta_{g\bar c  }^{-(5,1)}.
\end{align}
We use \textsc{PolyLogTools}~\cite{Duhr:2019tlz} and {\tt{MCFM-10.3}}~\cite{Campbell:2015qma,Campbell:2019dru} to obtain the various contributions.

For NNLO CCDY the VFNS scheme is validated and shown in table~\ref{Table:Fiducial-CDYas21} and \ref{Table:Fiducial-CDYas22}  for $W^{\pm}$ respectively. The $\Sigma_{\ln[m],[f]}$ and  $\Sigma_{n_f,[f]}$ is the quantity obtained used eq.~\eqref{eq:Sigma_power_corrections}.
\begin{table}
\begin{center}
\renewcommand{\arraystretch}{1.5}
\begin{tabular}{|c|c|c|c|c|}
\hline
 $m_Q$(GeV)& $\Sigma^{(5,0,k)}_{M,[c]}$(nb) &   $\Sigma_{\ln[m],[c]}^{(5,0,k)}$ (nb) & $\Sigma_{n_f,[c]}^{(5,0,k)}$ (nb) & $\Sigma_{pc,[c]}^{(5,+1,2)}$(nb)                                      \\ \hline
  0.1 & $3.13786 \pm 0.88263\times 10^{-3}$ & $3.50868$       &       $\num{-0.379504362920831113}$ & 0.00868492 \\ \hline
  $\num{0.517947467923121}$ & $2.22798 \pm 0.16191\times 10^{-3}$ & $2.59843$       &       $\num{-0.379504362920831113}$ & 0.00905337        \\ \hline
1.4 & $1.69405 \pm 0.16664\times 10^{-3}$ & $2.06338$       &       $\num{-0.379504362920831113}$ & 0.0101773        \\ \hline
\end{tabular}
\end{center}
	\caption{Predictions for the NNLO corrections to the CCDY ($W^{+}$) process for the $Q$-bin (50 GeV, 150 GeV). The contributions from the massive power-corrections at $\mathcal{O}(\alpha_s^2)$ are shown for the central scale for different $m_c$ values. }
\label{Table:Fiducial-CDYas21}
\end{table}

\begin{table}
\begin{center}
\renewcommand{\arraystretch}{1.5}
\begin{tabular}{|c|c|c|c|c|}
\hline
$m_Q$(GeV)& $\Sigma^{(3,0,k)}_{M,[c]}$(nb) &   $\Sigma_{\ln[m],[c]}^{(5,0,k)}$(nb) & $\Sigma_{n_f,[c]}^{(5,0,k)}$(nb) & $\Sigma_{pc,[c]}^{(5,-1,2)}$ (nb)                                     \\ \hline
$\num{0.372759372031494}$ & $2.58822 \pm 0.28766\times 10^{-3}$ & $2.98101$       &       $\num{-0.40152}$ &  0.00872798 \\ \hline
 1.4 & $1.83102 \pm 0.18157\times 10^{-3}$ & $2.2221$       &       $\num{-0.40152}$ & 0.0104385       \\ \hline
\end{tabular}
\end{center}
	\caption{Predictions for the NNLO corrections to the CCDY ($W^{-}$) process for the $Q$-bin (50 GeV, 150 GeV). The contributions from the massive power-corrections at $\mathcal{O}(\alpha_s^2)$ are shown for the central scale for different $m_c$ values. }
\label{Table:Fiducial-CDYas22}
\end{table}

\bibliography{literature} 
\bibliographystyle{utphysM}
\end{document}